\documentclass[fleqn,usenatbib]{mnras}

\pdfoutput=1

\newcommand{\noopsort}[1]{}

\usepackage[T1]{fontenc}
\usepackage{ae,aecompl}
\usepackage{threeparttable}
\usepackage{pdflscape}
\usepackage{rotating}

\usepackage{graphicx}	
\usepackage{amsmath}	
\usepackage{amssymb}	
\usepackage{xcolor}
\DeclareMathOperator{\Tr}{Tr}

\numberwithin{equation}{section}

\title[spatial density aggregation of dark matter]{Tensorial solution of the Poisson equation and the\\ 
dark matter amount and distribution of UGC 8490\\
and UGC 9753}

\author[P. Repetto]{
P. Repetto$^{1}$\thanks{E-mail: prsatch6@gmail.com}
\\
$^{1}$Via Gallaretta 42, Castelletto d'Orba, Alessandria, C.P. I-15060, Italia\\
}

\date{Accepted XXX. Received YYY; in original form ZZZ}

\pubyear{2015}

\begin{document}
\label{firstpage}
\pagerange{\pageref{firstpage}--\pageref{lastpage}}
\maketitle

\begin{abstract}
In the first part of this article we expand three fundamental aspects of the methodology connected to the determination of a relation
among the spatial density and the gravitational potential that can be specialised to distinct mass density agglomerations. As a consequence, we obtain
general relations for the diagonal entries of a square symmetric matrix without zeros, we provide an expression of the gravitational
potential, suitable, to represent several different mass density configurations, and we determine relations for the semi-axes of a triaxial spheroidal
mass distribution, as a function of the spheroid mass density, volume density and radius. In the second part of this manuscript, we 
employ the tools developed in the first part, to analyse the mass density content and the inner and global structure of the dark matter haloes of 
\mbox{UGC 8490} and UGC 9753, through the fits to the dark matter rotation curves of the two galaxies, assuming a triaxial spheroidal dark matter mass
configuration. We employ the Navarro Frenk and White, Burkert, DiCintio, Einasto and Stadel dark matter models, and we obtain that both a
cored Burkert and cuspy DiCintio and Navarro Frenk and White inward dark matter distributions could represent equally well the observed data, furthermore we
determine an oblate spheroidal dark matter mass density configuration for UGC 8490 and UGC 9753. The latter outcome is confirmed by the estimation of the
gravitational torques exerted by the dark matter halo of each analysed galaxy, on the corresponding baryonic components.
\end{abstract}




\section{Introduction}\label{sec:s1}

The global quantity and configuration of mass in galaxies represents an interesting astrophysical problem of many
facets, that was first addressed in the period among the forties and the sixties, by several authors, in an attempt to determine, on the one hand, an
approximate value of the whole mass of the studied galactic systems, and, on the other hand, the more adequate volumetric aggregation
of the corresponding galactic masses up to a certain radial extent, defined by the utmost radius of the observed rotation curves (RCs), of each analysed
object. In that regard, the most noteworthy contributions, are those of \citet{Burbidge1959} and~\citet{Brandt1960}, that determine the total mass
distribution of some nearby galaxies, considering that the observable part of those galaxies could be approximated by a set of concentric
oblate spheroids. 

The discrepancy among the total mass and the luminous mass of a certain number of local galaxies was determined by some authors, including
\citet{Hubble1929}, ~\citet{Babcock1939}, ~\citet{Wyse1942}, through a kinematic and photometric analysis, together with a
study of the RCs of the selected galaxies, to obtain an estimation of the total dynamic mass of the analysed galactic systems. The first
studies, regardless of their fundamental importance, are unable to establish unambiguously the correct value of the mass-to-light (M/L) ratios of the
analysed objects, principally due to the large uncertainties associated to the spectroscopic and photometric measurements necessary to obtain an estimation
of the luminosity and total mass of the selected galaxies. 

The determination of more robust M/L ratios, together with other important results, were accomplished at the end of the seventies, as a
consequence of the rapid advancement of the spectroscopic and photometric observational techniques utilised for the data collection, as well as, the
improvement of the procedures for the reduction and analysis of the data. Some of the more prominent works that illustrate the fundamental changes
delineated above, are outlined in the following. \citet{Rubin1978} studied a sample of 10 high surface brightness massive spirals, of morphological types
from Sa to Sc, to ascertain, among other findings, an average optical blue band M/L ratio of 3.5. In a similar work, \citet{Rubin1985}, using optical
luminosities in the blue band, determine, jointly with other outcomes, the mean values of the M/L ratios, of a sample of 54 spiral galaxies to be 6.2, 4.2 and 2.6, for the morphological types Sa, Sb, Sc, respectively, within the isophotal radius 25. \citet{Burstein1985}, and~\citet{Burstein1986} analysed a
total sample of 80 RCs of Sa, Sb and Sc galaxies, determining the whole gravitating mass of these systems. The quoted articles, together with several other
works \citep[e.g.][]{Kent1987, Persic1990, Forbes1992, Persic1996, Takamiya2000}, established stronger observational evidences, on galactic scale, of the
existence of a non observable mass component, specifically the dark matter (DM), whose total amount and related kinematical and dynamical effects could
eventually affect the evolution of any other galactic constituent, as well as of the entire galaxy.
 
From the end of the nineties to the present time, the majority of the observational efforts have been directed to improve the quality of the observed RCs,
ameliorating the spatial and spectral resolution of the photometric and spectroscopic data, to determine much more reliable constraints on the estimated
baryonic and DM masses, and to establish the actual total mass distribution of galaxies of different morphological types, based on more advanced and robust
methods of data reduction and analysis. In particular the principal interests are devoted to the study of the inner mass configuration of dwarf galaxies of
low surface brightness, where it is supposed that DM rules the entire dynamics, and for that reason its gravitational effects on the other galactic
components, could be studied in a more precise and direct manner \citep[e.g.][]{Blais2001, Blais2004, Spano2008, deBlok2008, Kuzio2009}. The contrast among
the theoretical predictions of an infinite amount of mass in the internal regions of galaxies and the observational detections of a finite DM inner mass
distribution, represents the cuspy/core issue, addressed for several years by many authors, from a theoretic and observational perspective, and that still
constitutes an unsettled matter of large study and debate \citep[e.g.][]{Salucci2010, Plana2010, Karukes2015, Allaert2017, Korsaga2018, deBlok2018}. 

The current manuscript is dedicated to the creation of some general tool to analyse the dynamics and kinematics of various 3D mass
distributions, and, as an application, of the novel developed technique, we perform the fits to the H$\alpha$ and HI DM RCs of UGC 8490 and UGC 9753, to obtain the total and DM mass of both galaxies, together with its global spatial configuration. In the following we proceed illustrating the principal matters studied in this article.

In the first part of the present work we propose a generalisation of some of the most important aspects of the methodology introduced in \citet{Repetto2018}
(RP18), in particular, we diagonalize a square symmetric matrix with three columns and rows (3SSM) without null elements, we determine an
integral solution of the Poisson equation appropriate to describe the gravitational potential of several different 3D mass density distributions, and we
find out the relations of the semi-axes of a hypothetical DM triaxial spheroid, as parametric ratios of the DM masses, scale radii and the corresponding
volume densities. The second part of the current paper examines the potentiality of the extension of the approach of RP18, performed in the first part of the current study, investigating the DM and total mass content of UGC 8490 and UGC 9753, hypothesising that the DM haloes of
both galaxies could be appropriately described through a triaxial spheroidal mass arrangement of non uniform volume density. In particular we employ the DM
density profiles of Navarro, Frenk and White, \citet{Navarro1996} (NFW), Burkert, \citet{Burkert1995} (BKT), DiCintio, \citet{DiCintio2014} (DCN),
Einasto, \citet{Einasto1965} (EIN) and Stadel, \citet{Stadel2009} (STD). We determine that the inner DM mass distribution of UGC 8490 and
\mbox{UGC 9753} analysed through the H$\alpha$ and HI DM RCs are well represented by the following DM velocity profiles: the cored BKT and the cuspy DCN, for the
H$\alpha$ DM RC of UGC 8490 and UGC 9753, separately, and the cuspy DCN and NFW, for the HI DM RC of UGC 8490 and UGC 9753, respectively. The overall DM
distribution of the two galaxies analysed in this work is well described by a flattened DM mass density configuration. It is important to note that in this
manuscript we do not attempt to address the cuspy/core discrepancy, given the insufficient number of RCs analysed and the non adequate spatial resolution of
the HI DM RCs, instead we assess the potentiality of the determined integral solution of the Poisson equation, by means of the application
presented in section~(\ref{sec:s6}).

The content of the current manuscript is the following: in Section~(\ref{sec:s2}), and in the Appendix~(\ref{sec:a1}) we present some
relations to transform a non orthogonal system of curvilinear coordinates to an orthogonal one, in Section~(\ref{sec:s3}) and in the Addendum~(\ref{sec:a2})
we determine a general integral solution of the Poisson equation, in Section~(\ref{sec:s4}) and in the Supplement~(\ref{sec:a4}) we determine the semi-axes
relations for a triaxial spheroidal DM configuration, in Section~(\ref{sec:s5}) and Addendum~(\ref{sec:a3}) we specialise the Poisson integral solution to a
triaxial spheroidal mass distribution and characterise the rotation through an exact formula in the corresponding equatorial plane, in
Section~(\ref{sec:s6}) and the relative subsections we apply the previous results to the determination of the amount and distribution of the DM mass of 
UGC 8490 and UGC 9753, in Section~(\ref{sec:s7}) we determine the values of the semi-axes of the hypothesised DM triaxial spheroid by means of the
gravitational torques method, we present two mock data trials, to appraise the robustness of the conceived methodology, and a
comparison with other studies, in Section~(\ref{sec:s8}) is presented a discussion of the principal attainments of the current article, whereas the conclusions are
detailed in Section~(\ref{sec:s9}).

\section{Extension of the methods of RP18}\label{sec:s2}

In the next three sections, that constitute the first part of this work, we present a generalisation of the essential steps that are
inherent to the methodology developed in RP18, in addition we provide some relations to derive an orthogonal curvilinear coordinates system from a
non orthogonal one, and we determine the semi-axes of a triaxial spheroidal distribution of matter as a function of the spheroidal mass density, volume
density and radius. Specifically, the procedural extension is concerned with a general solution of the Poisson equation applicable to many
different mass configurations, the diagonalization of a 3SSM with non null elements, and the determination of three expressions for the semi-axes
of the supposed triaxial spheroidal mass distribution. The introduced methodology expansion is appropriate for 3D mass configurations, and the subsequent
astrophysical application, that represents the second part of this article, takes advantage of that extension.

\subsection{Diagonalization of a 3SSM without null entries}\label{sec:s21}

In this section we compute the diagonal elements of a 3SSM that has all entries different from zero. We solve the system of the
three equations derived through the equalities between the coefficients of the characteristic polynomial of the 3$\times$3 diagonal matrix and the
characteristic polynomial of the 3SSM without null elements. The most important details about the relevant equations and the calculations accomplished, 
are given in Appendix~(\ref{sec:a1}), whereas the nine solutions, are reported below. We indicate the 3SSM through the symbol $\mathrm{M_{ml}}$, its
corresponding elements with small capitals v letters, and the entries of the diagonal matrix with capital v letters.
The nine relations, that transform a non orthogonal system of curvilinear coordinates to an orthogonal one are expressed, in compact form, through the following equations:

\begin{equation}\label{eqn:eq1}
V^{(+,-)}_{ii}=\left\{\left[\frac{\Tr{(M_{ml})}\pm\Delta^{(i)}}{2}\right]\right\}
\end{equation}

\begin{equation}\label{eqn:eq2}
V^{(+,-)}_{kk}=\frac{1}{2}\left\{V^{(-,+)}_{ii}\pm\Delta^{(i)}_1\right\} 
\end{equation}

\begin{equation}\label{eqn:eq3}
V^{(-,+)}_{jj}=\frac{1}{2}\left\{V^{(-,+)}_{ii}\mp\Delta^{(i)}_1\right\}
\end{equation}
  
\noindent where the quantities $\Delta^{(i)}$ and $\Delta^{(i)}_1$ are defined according to the ensuing relations:

\begin{align}\label{eqn:eq4}
&\Delta^{(i)}=\sqrt{\left[v_{kk}+v_{jj}-v_{ii}\right]^2+4\left[v^2_{ik}+v^2_{ij}\right]}\nonumber\\
&\Delta^{(i)}_1=\sqrt{\left[V^{(-,+)}_{ii}\right]^2-4\left[v_{kk}v_{jj}-v^2_{kj}\right]}
\end{align}

\noindent The indices $\mathrm{i,k,j}$ satisfy the condition i $\neq$ k $\neq$ j and range from 1 to 3, through all possible permutations.
Particular instances of the solutions provided above, are determined whenever pairs of non diagonal matrix elements are zero. The solution of RP18 can be
obtained, for instance, setting $\mathrm{i=2,k=3,j=1}$, $\mathrm{v_{12}=v_{23}=0}$ and $\mathrm{v_{13}}$ $\neq$0. Other specific solutions are determined
interchanging $\mathrm{13}$ with $\mathrm{12}$ and $\mathrm{13}$ with $\mathrm{23}$, and considering $\mathrm{i=3,k=2,j=1}$, and $\mathrm{i=1,k=3,j=2}$,
respectively. The solutions above constitute a set of fundamental transformations among non orthogonal and orthogonal systems of curvilinear coordinates.
In the present manuscript we employ equation~(\ref{eqn:eq1}), ~(\ref{eqn:eq2}) and ~(\ref{eqn:eq3}) to orthogonalise the non orthogonal curvilinear system
of coordinates utilised to address the particular issue analysed in the subsequent part of this work.         

\section{Integral solution of the Poisson equation}\label{sec:s3}

In the present section we determine an integral solution of the Poisson equation, to obtain an expression of the gravitational potential,
appropriate to describe several different mass density configurations. In appendix~(\ref{sec:a2}) we provide the complete details of that derivation. The
gravitational potential formula, established in this work, expresses the radial derivative of the 3D gravitational potential as an integral function of the
diagonal entries of the metric tensor matrix, and therefore it is supposed that the spatial density points can be properly described by a system of
Cartesian or curvilinear coordinates that is orthogonal, or alternatively, that can be reduced to orthogonal form, through matrix diagonalization. It is
evident from addendum~(\ref{sec:a2}), that we actually obtain the gradient of the 3D gravitational potential, not only its radial derivative, however we
concentrate on that latter quantity solely for the purpose of the analysis performed in this article. The radial partial derivative of the gravitational
potential can be expressed in the following manner:

\begin{align}\label{eqn:eq5}
&\frac{\partial \phi_1(u_1,u_2,u_3)}{\partial u_1}=\nonumber\\
&=\sqrt{\frac{V_{11}}{V_{22}V_{33}}}\left[4\pi G \int_0^{u_1}\rho_1(v_1,v_2,v_3) \sqrt{V} dv_1 +C_1\right].
\end{align}
 
\noindent where $\mathrm{G}$ is the Newtonian constant of gravitation. The volume density is represented by the quantity 
$\mathrm{\rho_1(u_1,u_2,u_3)}$, and a non specific system of curvilinear coordinates $\mathrm{(u_1,u_2,u_3)}$ defines the loci of the spatial density
points. The solution~(\ref{eqn:eq5}), encompasses both the Laplace and Poisson equations, i.e. the homogeneous and non homogeneous differential equation.
The quantity $\mathrm{V}$ corresponds to the determinant of the diagonal metric tensor matrix, and the constant $\mathrm{C_1}$ have to be expressed in units
of $\mathrm{kpc\,km^2 s^{-2}}$. The particular problem studied in this manuscript requires the application of the solution~(\ref{eqn:eq5}) to a spheroidal
system of curvilinear coordinates and, in addition, for the specific objective of this research, we have to omit the homogeneous part of that solution.

\section{Triaxial spheroid semi-axes expressions}\label{sec:s4}

This section illustrates the relations obtained for the semi-axes of a triaxial spheroidal mass density configuration, that can be
parametrized through the resultant DM fit parameters, such as, for instance, the DM halo mass and scale radius, and the corresponding DM halo volume
density. In the subsequent part of the present section we report the relevant equations, whereas the complete derivation of the semi-axes parametrizations
is left to appendix~(\ref{sec:a4}). We determine, as a first step, the products of the semi-axes, i.e. the quantities $\mathrm{a_1a_2}$, $\mathrm{a_1a_3}$
and $\mathrm{a_2a_3}$, considering the relations of the 3D total spheroidal mass density, and the related 2D surface mass densities on the equatorial
ellipse, and the other two perpendicular ellipses, within the triaxial spheroid. Once obtained the semi-axes products, we determine the semi-axes
$\mathrm{a_1}$, $\mathrm{a_2}$ and $\mathrm{a_3}$, for a triaxial spheroidal mass density distribution, and in the case where the equatorial ellipse of
the analysed spheroidal density aggregation is a circle. The equations of the semi-axes in the case $\mathrm{a_1 \neq a_2 \neq a_3}$ are the following:  

\begin{align}\label{eqn:eq6}
&a_1 = \left[\frac{1}{4\pi}\left(\frac{I_{1d}}{I_{1M}}\right)\left(\frac{I_{2M}}{I_{2d}}\right)\left(\frac{I_{3M}}{I_{3d}}\right)\right]^{\frac{1}{2}}\nonumber\\
&a_2 = \left[\frac{1}{4\pi}\left(\frac{I_{1M}}{I_{1d}}\right)\left(\frac{I_{3M}}{I_{3d}}\right)\left(\frac{I_{2d}}{I_{2M}}\right)\right]^{\frac{1}{2}}\nonumber\\
&a_3 = \left[\frac{1}{4\pi}\left(\frac{I_{1M}}{I_{1d}}\right)\left(\frac{I_{2M}}{I_{2d}}\right)\left(\frac{I_{3d}}{I_{3M}}\right)\right]^{\frac{1}{2}}
\end{align}

\noindent where the quotients $\mathrm{\frac{I_{id}}{I_{iM}}}$ and $\mathrm{\frac{I_{iM}}{I_{id}}}$, can be expressed as constant ratios of
the resultants DM halo masses and scale radii, and the correspondent volume densities, and their reciprocal, respectively, as delineated in the addendum~(\ref{sec:a4}). The semi-axes relations for the instance $\mathrm{a_1=a_2 \neq a_3}$, are given through the ensuing formulas:

\begin{align}\label{eqn:eq7}
&a_1=a_2=\left[\frac{1}{4\pi}\left(\frac{I_{3M}}{I_{3d}}\right)\right]^{\frac{1}{2}}\,a_3=\left[\frac{1}{4\pi}\left(\frac{I_{3d}}{I_{3M}}\right)\right]^{\frac{1}{2}}\left[\frac{I_{1M}}{I_{1d}}\right]\nonumber\\
&a_3 = \left[\frac{1}{4\pi}\left(\frac{I_{3d}}{I_{3M}}\right)\right]^{\frac{1}{2}}\left[\frac{I_{2M}}{I_{2d}}\right]
\end{align}

\noindent The products of the semi-axes $\mathrm{a_1a_2}$, $\mathrm{a_1a_3}$ and $\mathrm{a_2a_3}$, are obtained in appendix~(\ref{sec:a4}), and the relations~(\ref{eqn:eq6}), and~(\ref{eqn:eq7}), are determined solving the system of equations of the semi-axes products. In the current manuscript we are concerned with a specific issue, and we employ the semi-axes expressions given by equation~(\ref{eqn:eq6}), because we suppose that the spheroidal DM mass distribution analysed is well represented by such type of spheroid. The latter assumption is verified a posteriori by
means of the DM fits and the determined DM halo resultant parameters. It is important to note that the results of this section as well as of the
addendum~(\ref{sec:a4}), are general and independent of the particular matter studied in the present work, and therefore their applicability encompass a
much more ample spectrum, than the simple example considered in the current analysis.

\section{Rotation formula in the equatorial plane}\label{sec:s5}

In the present section we outline the region of the spheroidal mass distribution, where rotational motions should exist, and for that
zone we determine an exact rotation formula, derived from the solution of the Poisson equation, obtained in section~(\ref{sec:s3}). The latter solution have
to be specialised to a spheroidal system of orthogonal curvilinear coordinates, employing the diagonal solution of section~(\ref{sec:s21}), that transforms
a non orthogonal system of curvilinear coordinates to orthogonal form. The triaxial spheroidal system of curvilinear coordinates we have selected is not
orthogonal in its original configuration, and therefore it is necessary to apply the diagonalization relations determined in section~(\ref{sec:s21}), to
achieve its orthogonality. The diagonalization process employs the non diagonal expressions of the metric tensor matrix elements obtained in appendix~(\ref{sec:a3}), because the diagonalization formulas of section~(\ref{sec:s21}) are expressed as functions of the non diagonal metric tensor matrix
entries. The relations of addendum~(\ref{sec:a3}), before to be inserted into the diagonalization expressions of section~(\ref{sec:s21}), have to be
evaluated for certain azimuthal and zenithal angles, depending on the regions where the rotational motion could exist, and the precise details of that
procedure together with the resultant relations are given in the subsequent part of this section. 

In the equatorial plane (EP), the components of the angular momentum of a triaxial spheroid of non uniform density with respect to the 
$\mathrm{x_1}$ and $\mathrm{x_2}$ axes vanish, whereas the angular momentum component along the $\mathrm{x_3}$ axis is different from zero, and therefore
the rotation is parallel to the EP. In the meridional plane as well as in the corresponding perpendicular plane, the main rotation pattern is about the 
$\mathrm{x_2}$ and $\mathrm{x_1}$ axes, respectively, and as a consequence the rotation is perpendicular to both planes. We particularise the velocity
relations in the EP, because the RC is usually build from radial velocities on the sky, subtracted of the systemic, and de-projected to rotation and other
possible components in the EP. We are aware that in the EP, the rotation coexists together with other types of motions, nevertheless the EP kinematics
resembles much more than other planes the velocity field kinematics of an observed galaxy, for what noticed above about the angular momentum components, 
and consequently the velocity formula have to be computed in the EP. The latter evidence is also confirmed by theoretical studies of the evolution of
stellar orbits into a non spherical gravitational potential well \citep{Touma1997}. The starting point, for the computation of the velocity formula, is
represented by the relation~(\ref{eqn:eq5}), the non orthogonal entries of the metric tensor matrix specialised for the azimuthal and zenithal angles of
the EP, the elements of the orthogonal metric tensor matrix, calculated through the expressions~(\ref{eqn:eq1}),~(\ref{eqn:eq2}), and~(\ref{eqn:eq3}), as
functions of the non orthogonal metric tensor matrix elements, and the DM haloes density profiles to be integrated to obtain the velocity relations
according to the prescription of equation~(\ref{eqn:eq5}). The resultant equation that describes the rotational motion in the EP is reported below:

\begin{align}\label{eqn:eq8}
&\frac{\partial \phi_1(v_1,0,\frac{\pi}{2})}{\partial v_1}=\sqrt{\frac{V_{11}}{V_{22}V_{33}}}\left[G M_{DM}(v_1)+C_1\right]\nonumber\\
\end{align}

\noindent where the quantity $\mathrm{M_{DM}(v_1)}$ is the DM mass density, that can be determined for each DM halo through the integration
of the DM haloes density profiles. The RC of each DM halo, is measured on segments that are parallel to the semi-minor axis, and the latter fact explains the selection of the azimuthal angle $\mathrm{v_2=0}$. The choice of the EP sets the zenithal angle $v_3=\frac{\pi}{2}$. The square root containing the term
of the diagonal metric tensor matrix is computable from the relations of section~(\ref{sec:s21}), and the resultant DM models RCs, considering only the
Poisson part of the solution, i.e. $\mathrm{C_1=0}$, are given by the following relation:

\begin{equation}\label{eqn:eq9}
v_1\frac{\partial \phi_1(v_1,0,\frac{\pi}{2})}{\partial v_1}=\frac{G M_{DM}(v_1)}{v_1}\frac{a_1}{a_2a_3}
\end{equation}

\noindent The triaxial spheroid semi-axes $\mathrm{a_1,a_2,a_3}$ that enter into equation~(\ref{eqn:eq9}), are parametrized according to
the recipe disclosed in section~(\ref{sec:s4}), specifically are expressed as quotients of the initial fitting parameters, such as the initial DM halo masses, scale radii, and the derived initial volume densities. It is important to clarify that the DM halo masses and scale radii represent the solely free
parameters of the fitting process of the DM RCs of UGC 8490 and UGC 9753, and the triaxial spheroid semi-axes are connected to the variations of those two
free parameters through the relations given in section~(\ref{sec:s4}), and determined in appendix~(\ref{sec:a4}). The optimal fitting DM halo masses and scale radii, obtained as a result of the DM RCs fitting process of UGC 8490 and UGC 9753, determine the values of the DM haloes semi-axes, to unveil the actual spatial configuration of the DM mass density of both galaxies. We employ equation~(\ref{eqn:eq9}), to perform the fit of the DM RCs of UGC 8490 and UGC 9753. The details of the fitting procedure, and the resultant total and DM masses of UGC 8490 and UGC 9753, as well as, the overall DM mass
distribution, for each DM model considered, are given in the next section.

\section{Application to the DM RC\lowercase{s} fit of \mbox{UGC 8490} and UGC 9753}\label{sec:s6}

The resultant DM halo mass and scale radius of every DM velocity profile considered in this study, are of great importance to discriminate among the sort of
inner mass aggregation, and the spherical and oblate/prolate triaxial spheroidal solutions. In the subsequent parts of the present section we delineate
primarily the Fourier analysis of the velocity fields of both galaxies to validate the hypothesis that the gaseous components are moving on
nearly circular orbits, the fitting methodology, the DM models utilised, the necessary restrictions on the triaxial spheroid semi-axes, and the principal
results attained. In the following we provide in tabular form the triaxial spheroidal DM masses and scale radii, whereas the best DM models fits are
reported in graphical form. In this section we take advantage of the DM RCs of UGC 8490 and UGC 9753, determined by RP18, all the details about the
derivation of both DM RCs, together with other important related things can be found in RP18. The current research could be considered as an improvement and
extension of some fundamental aspects of the analysis of RP18.

\subsection{Fourier analysis of the velocity fields of \mbox{UGC 8490} and UGC 9753}\label{sec:s61}

We apply the methodology of \citet{Schoenmakers1997} (Sh97) to accomplish a Fourier harmonic analysis of the HI radial velocity fields of UGC 8490 and UGC 9753. The Sh97 technique is available through the reswri package, that forms part of the Groningen Image Processing System (GIPSY) \citep{van1992,vog2001}. The strategy of Sh97, measures the deformation induced by a triaxial non spherical perturbation to the original axisymmetric gravitational potential, representing the non axisymmetric perturbation as a sum of Fourier harmonic components. The harmonic fit to the HI velocity fields of \mbox{UGC 8490} and UGC 9753, should detect possible deviation from axisymmetry in the gravitational potential of both galaxies, to determine the amplitude of the radial motions associated with the kinematics of the gaseous components, and the corresponding elongation of the resultant gravitational potentials. Adopting the notation of Sh97, we denote with $\epsilon_{\mathrm{pot}}$ the elongation of the global gravitational potential, and with 
$\varphi_2$, an angular quantity, that is related to the azimuthal angle and the observer angle of sight.

\begin{figure*}
\centering
\includegraphics[width=1.0\hsize]{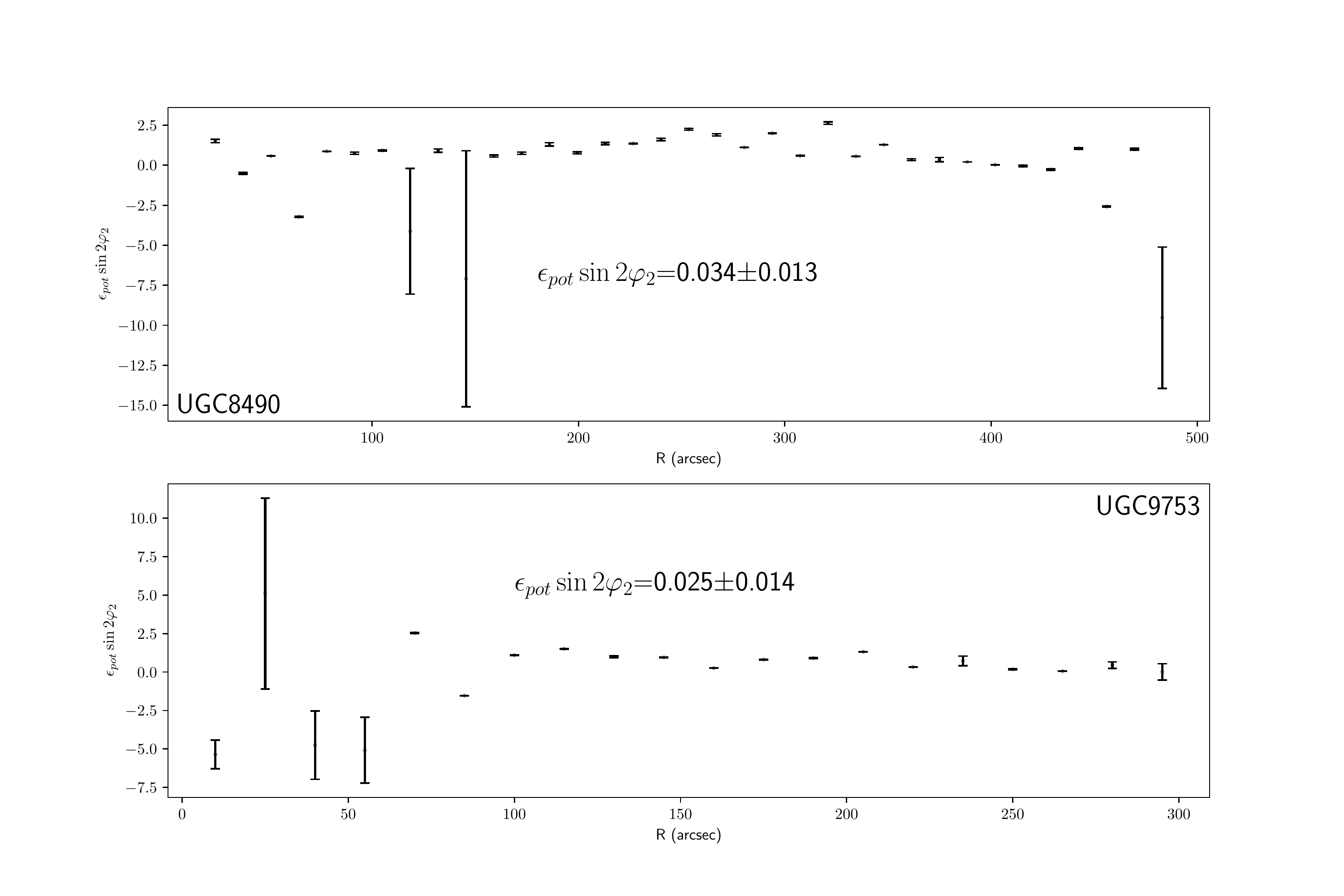}
\caption[f1.pdf]{Gravitational potential elongation as a function of radius for UGC 8490 and UGC 9753. The exact values of the elongation together with the respective errors are showed within the figures. The elongation values are less than 0.1, and therefore we conclude that the gravitational potential is globally axisymmetric and that the gas orbits are well approximated by circles.}
~\label{fig1}
\end{figure*}

The accomplished harmonic analysis of the HI velocity fields of UGC 8490 and UGC9753 ascertains that their average gravitational potential elongation is 0.034$\pm$0.013, and 0.025$\pm$0.014, respectively, and therefore the approximation of nearly circular orbits for the HI gas components of both galaxies is adequate enough to describe the global motion of the neutral hydrogen of UGC 8490 and UGC 9753. The corresponding radial plots of the gravitational potentials elongations of both galaxies are displayed in figure~(\ref{fig1}).

\subsection{Fitting strategy and semi-axes settings}\label{sec:s62}

The fits to the HI and H$\alpha$ DM RCs of UGC 8490 and \mbox{UGC 9753} were performed through the {\bf minuit} fitting routine, that is part of the ROOT
package \citep{ReneBrun}, a versatile data analysis tool created at CERN. We made use of five DM models, expressly, the NFW, BKT, DCN, EIN, and STD DM
haloes.

In the present work we prefer to fit separately the H$\alpha$ and HI RCs, because as detailed in section~(\ref{sec:s72}), the HI disc of both galaxies displays a very perturbed dynamics, in contrast, the H$\alpha$ disc of UGC 8490 and UGC 9753 exhibits a much more quiet dynamical behaviour, and therefore we argue that the two gaseous tracers should not necessarily describe exactly the same gravitational potential, and as a consequence we perform the fits of the H$\alpha$ and HI DM RCs. 

The adopted fitting methodology can vary an indefinite number of parameters, nonetheless the DM models of NFW, BKT and DCN are characterised by two fundamental quantities, the DM halo mass and scale radius, whereas the DM halos of EIN and STD present an additional parameter that controls the spatial mass density arrangement of the DM halo, and as a consequence, for the EIN and STD DM halos there are 
three free parameters. The resultant EIN shape parameter converged to the values of 2.3 and 3.9, 5.0 and 1.4, 
for the triaxial spheroidal fits to the H$\alpha$ and HI DM RCs of UGC 8490 and UGC 9753, separately, whereas the final STD $\lambda$ parameter converged
to the values of 3.7 and 4.4, 5.7 and 3.5, for the triaxial spheroidal fits to the H$\alpha$
and HI DM RCs of UGC 8490 and UGC 9753, respectively. We have to clarify that the values of the EIN form parameters, determined through our fitting analysis, are in total agreement with the outcomes of the Cold DM N-body numerical simulations of \citet{Navarro2004}, and \citet{Dutton2014} (DM14), given that, the EIN density profile, that those authors employ has the reciprocal shape index with respect to our profile, as also observed by \citet{Chemin2011} (CH11). The STD $\lambda$ index is also concordant with the prescription of STD, given that in the velocity formula derived in \citet{Repetto2015}, the STD index appear as the reciprocal of the STD index $\lambda$. We found different values of the indexes parameters of EIN and STD, for the H$\alpha$ and HI RCs of \mbox{UGC 8490} and \mbox{UGC 9753}, and a possible interpretation could be that the two distinct tracers denote different
parts of the same gravitational potential, as partially confirmed by the fact that the resulting H$\alpha$ and HI DM masses and scale radii delineate a
dissimilar DM mass density distribution among the H$\alpha$ and HI gaseous tracers. 

The range of variation of the DM haloes mass and radius is $[10^8,10^{13}]$ M$_{\odot}$ and $[0.1, 500.0]$ kpc, respectively. 

The fitting procedure have two important indicators to decide if a given
solution can be considered satisfactory, the reduced $\chi^2_r$, that are $\chi^2$ normalised for the number of degrees of freedom, and the estimated
distance to the minimum (EDM). In the current work we reformulate the $\chi^2$ definition, adopting the square root of the observed DM RCs
data as the divisor of the differences between the DM models and the DM RCs data, and the resultant $\chi^2_r$ values seem more concordant
with the quality of the presented DM RCs fits. The EDM values are determined on the base of a Monte Carlo search, associated to a Metropolis minimization
algorithm, that repeatedly measures the minimum distance among all the solutions found through the fitting process, to ascertain the best possible
minimization result.

The definition of the oblate and prolate spheroids employed in the present work, relies on the values of the
semi-axes $\mathrm{a_3}$, and it is dependent on the initial values of the DM fitting masses, scale radii, and corresponding spatial densities, as well as, their evolution during the fitting process, as we explain in more detail below. The initial values of the DM masses and scale radii of every DM halo, produce, at first, a triaxial DM configuration, that evolves according to their variation. The details about the triaxial spheroid semi-axes prescriptions, as a parametrization of the DM models masses and scale radii, are outlined in the following. As already mentioned in section~(\ref{sec:s4}), the semi-axes of the supposed triaxial spheroid are parametrized by the DM haloes masses, scale radii, and corresponding volume densities, that originate from the fitting process to the DM RCs of \mbox{UGC 8490} and UGC 9753. The integral quotients that enter in the definition of the triaxial spheroid semi-axes, delineated by
equations~(\ref{eqn:eq6}), and~(\ref{eqn:eq7}), can be expressed as adimensional ratios of the resulting fitting parameters, according to the following
prescription:

\begin{align}\label{eqn:eq10}
&\frac{I_{xM}}{I_{xd}}=\alpha_x \frac{M}{\rho v^3} & \frac{I_{xd}}{I_{xM}}= \frac{1}{\alpha_x}\frac{\rho v^3}{M}
\end{align}

\noindent where the constant $\mathrm{\alpha}$ varies depending on the DM models and other factors, explained below. The differences, in
the values of the constant $\mathrm{\alpha}$ for distinct DM models, are due to the dissimilar constants that define the five DM haloes density profiles
analysed in this work. The subscript $\mathrm{x}$ indicates the diverse magnitudes of the integral ratios depending on the three distinct mass density
symmetries considered within the triaxial spheroid, and defined through the three principal axes of symmetry, respectively. The relations~(\ref{eqn:eq10})
represent a compact form to express the integrals ratios, in the actual relations that we employed in the fit of the DM RCs of UGC 8490 and UGC 9753, the
constants $\mathrm{\alpha}$, are known real numbers for every DM halo studied, and their exact numerical values also comprehend the differences of amplitudes originated by the distinct volume density symmetries selected to determine the semi-axes products (e.g. appendix~(\ref{sec:a4})). The resultant
semi-axes values of each DM fit, are reported in table~(\ref{tab:tb1}), together with the corresponding DM masses, scale radii, and derived volume densities.
We employ as fitting formula the relation~(\ref{eqn:eq9}), specialised to the particular mass distribution of each DM halo, and with semi-axes parametrized
according to the prescriptions delineated above and in section~(\ref{sec:s4}). As in RP18 we adopt the approximation of DM14 for the
concentration parameter of the NFW DM halo, the DCN circular velocity is that of \citet{Kravtsov1998} (KT98), and we employed the approximations of
\citet{Ciotti1999} and of \citet{MacArthur2003} for the quantity $d_{\alpha}$ of the EIN DM model, in their respective regimes of validity.
 
\subsection{Best fitting models and resultant DM masses, scale radii and semi-axes}\label{sec:s63}

The description of the most important results include some necessary specifications about the best DM models, based on the $\chi^2_r$, EDM values, visual
quality of the fits and the number of points of the $\chi^2$ contours plots, since all those quantities together, establish the reliability of a given
solution, for every DM model studied in this manuscript. In addition, we present graphically the best fitting DM haloes, and we enumerate in four tables the totality of the fitting results. 

In the verbal account of the principal results, we emphasize two fundamental distinctive
attribute that every best fitting solutions should have, mainly depending on the intrinsic properties of the selected DM halo profiles, namely, the cuspy or
core behaviour. The latter difference is of great relevance to determine, as accurate as possible, the inner DM density profile, and as a consequence to
know whether the observed DM mass in the interiors of the studied galaxies has a finite or infinite amount. The further distinction among the spherical, oblate or prolate spheroidal character of the best solutions is also of great importance to determine the overall most probable DM mass distribution for UGC 8490 and UGC 9753. 

\begin{figure*}
\centering
\includegraphics[width=1.0\hsize]{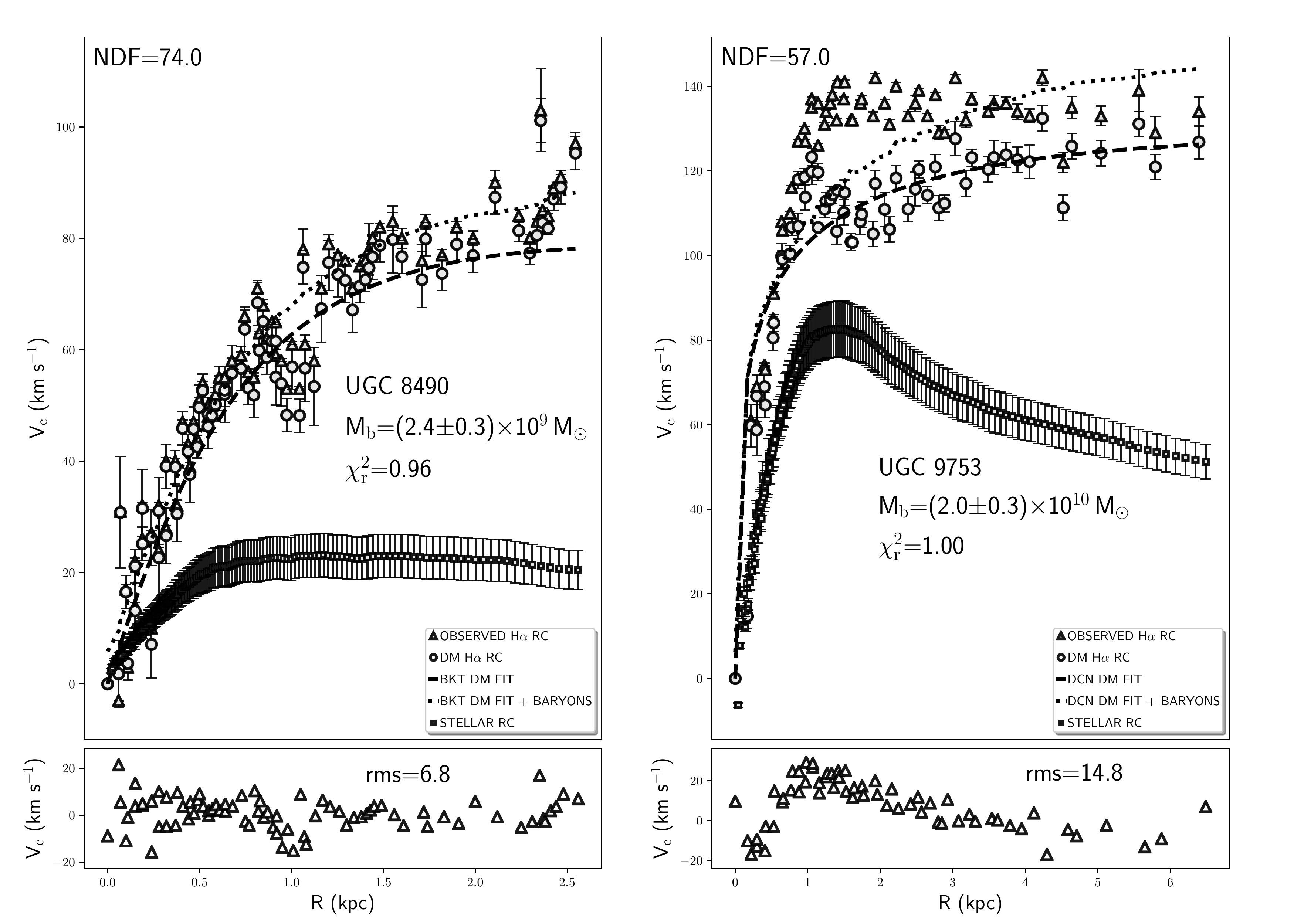}
\caption[f2.pdf]{{\it{Left}}: BKT DM fit (dashed line) to the H$\alpha$ DM RC of UGC 8490. The observed H$\alpha$ RC is displayed
together with the H$\alpha$ DM RC of the same galaxy. The dotted line denotes the BKT DM fit to the H$\alpha$ DM RC plus the stellar RC. The residuals in
the lower panel represent the subtraction of the squared BKT DM fit to the H$\alpha$ DM RC plus the squared stellar RC, from the squared observed H$\alpha$
RC. {\it{Right}}: Identical information for the DCN DM fit to the H$\alpha$ DM RC of UGC 9753.}
~\label{fig2}
\end{figure*}

The total baryonic masses of UGC 8490 and \mbox{UGC 9753} are estimated according to the methodology explained in RP18 and their respective values and
deviations are \mbox{(2.4$\pm$0.3) $\times$10$^9$ M$_{\odot}$} and \mbox{(2.0$\pm$0.3)$\times$10$^{10}$ M$_{\odot}$}, separately. The total stellar and gas masses of UGC 8490 and \mbox{UGC 9753} are $\approx$ 2.8$\times$10$^8$ M$_{\odot}$ and 2.2$\times$10$^9$ M$_{\odot}$, and $\approx$ 3.6$\times$10$^9$ M$_{\odot}$ and 1.6$\times$10$^{10}$ M$_{\odot}$, respectively. The total mass of both
galaxies is obtained by the sum of the total baryonic mass and the determined DM masses for each DM models adjusted. We refer, the interested reader, to
sec. 4 of RP18 for the methodology employed to obtain the stellar RCs, the HI+He+metals RCs, the DM RCs of 
both galaxies, that are displayed in Fig.~\ref{fig2}, and Fig.~\ref{fig4}, and all the relevant details concerning for instance the M/L
ratios. As already explained in RP18, we obtain the stellar RCs of UGC 8490 and UGC 9753, from stellar population synthesis studies, and we take into account the corresponding uncertainties once we build the 2D stellar maps, and then the stellar RCs. The HI+He+metals RCs are determined through the 2D HI total density maps, together with the relative errors, connected to distances, inclinations and position angles, as for the stellar RCs. The DM RCs of both galaxies are obtained subtracting the stellar RCs from the observed
H$\alpha$ RCs, and the stellar and HI RCs from the observed HI RCs. In our methodology, we do not consider the stellar and gaseous parts of a galaxy as parametric curves, instead, we prefer to obtain the baryonic portions of our galaxies from photometric and spectroscopic measurements of the respective
stellar and gaseous components.

The fitting process of the H$\alpha$ and HI DM RCs of \mbox{UGC 8490} and UGC 9753, determines that the overall DM distribution of both galaxies is adequately represented by oblate spheroids, whose semi-axes have different values depending on the galaxy and the DM model considered. The
corresponding semi-axes, together with the respective uncertainties are tabulated together with all the other results in Tables~\ref{tab:tb1}, ~\ref{tab:tb2}, ~\ref{tab:tb3}, and~\ref{tab:tb4}.

\begin{table*}
\centering
\caption{DM and total mass of UGC 8490: H$\alpha$ RC}
\label{tab:tb1}
\begin{threeparttable}
\begin{tabular}{p{1.2cm}p{1.2cm}p{1.2cm}p{1.2cm}p{1.2cm}p{1.2cm}p{1.2cm}p{1.2cm}p{1.2cm}}
\hline
DMHs\tnote{(1)} & M$_\mathrm{T}$\tnote{(2)}  & M$_{\mathrm{DM}}$\tnote{(3)}  
& R$_\mathrm{h}$\tnote{(4)} & $\mathrm{a_1}$\tnote{(5)} & $\mathrm{a_2}$\tnote{(6)} & 
$\mathrm{a_3}$\tnote{(7)} & $\chi^2_\mathrm{r}$\tnote{(8)} & EDM\tnote{(9)}\\
\hline
NFW &  11.7$\pm$2.8 & 11.5$\pm$2.8 & 231.5$\pm$22.2 & 40.2$\pm$2.4 & 40.2$\pm$2.4 & 3.2$\pm$0.2 & 0.7 & 2.3$\times$10$^{-8}$\\
BKT &  0.2$\pm$0.03 & 0.04$\pm$0.003 & 1.6$\pm$0.06 & 23.0$\pm$0.6 & 23.0$\pm$0.6 & 1.8$\pm$0.05 & 0.96 & 5.0$\times$10$^{-7}$\\
DCN &  0.9$\pm$0.05 & 0.7$\pm$0.02 & 13.0$\pm$0.1 & 104.1$\pm$0.6 & 104.1$\pm$0.6 & 8.3$\pm$0.05 & 0.6 & 1.6$\times$10$^{-10}$\\
EIN &  0.3$\pm$0.04 & 0.05$\pm$0.007 & 9.0$\pm$0.6 & 14.0$\pm$0.6 & 14.0$\pm$0.6 & 1.1$\pm$0.05 & 0.8 & 2.2$\times$10$^{-7}$\\
STD &  0.8$\pm$0.1 & 0.6$\pm$0.07 & 8.6$\pm$0.5 & 16.0$\pm$0.6 & 16.0$\pm$0.6 & 1.3$\pm$0.05 & 0.74 & 1.2$\times$10$^{-9}$\\
\hline
\end{tabular}
\begin{tablenotes}
\item[1] DM haloes.
\item[2] Total mass in units of 10$^{10}$ (M$_{\odot}$).
\item[3] DM mass in units of 10$^{10}$ (M$_{\odot}$).
\item[4] DM scale radius (kpc).
\item[5] DM halo major axis.
\item[6] DM halo axis orthogonal to the major axis on the EP.
\item[7] DM halo axis orthogonal to the EP.
\item[8] Reduced $\chi^2$.
\item[9] Estimated distance to the minimum.
\end{tablenotes}
\end{threeparttable}
\end{table*}

\begin{table*}
\centering
\caption{DM and total mass of UGC 9753: H$\alpha$ RC}
\label{tab:tb2}
\begin{threeparttable}
\begin{tabular}{p{1.2cm}p{1.2cm}p{1.2cm}p{1.2cm}p{1.2cm}p{1.2cm}p{1.2cm}p{1.2cm}p{1.2cm}}
\hline
DMHs & M$_\mathrm{T}$  & M$_{\mathrm{DM}}$  
& R$_\mathrm{h}$ & $\mathrm{a_1}$ & $\mathrm{a_2}$ & 
$\mathrm{a_3}$ & $\chi^2_\mathrm{r}$ & EDM\\
\hline
NFW &  2.5$\pm$0.3 & 0.5$\pm$0.03 & 57.0$\pm$1.7 & 212.0$\pm$4.1 & 212.0$\pm$4.1 & 17.0$\pm$0.3 & 0.82 & 1.7$\times$10$^{-9}$\\
BKT &  2.1$\pm$0.3 & 0.06$\pm$0.004 & 1.2$\pm$0.03 & 34.0$\pm$0.6 & 34.0$\pm$0.6 & 2.7$\pm$0.05 & 6.6 & 2.0$\times$10$^{-8}$\\
DCN &  8.0$\pm$0.4 & 6.0$\pm$0.05 & 23.0$\pm$0.2 & 52.0$\pm$0.6 & 52.0$\pm$0.6 & 4.1$\pm$0.05 & 1.0 & 9.9$\times$10$^{-9}$\\
EIN &  2.1$\pm$0.3 & 0.08$\pm$0.008 & 12.0$\pm$0.5 & 21.0$\pm$0.6 & 21.0$\pm$0.6 & 1.6$\pm$0.05 & 1.6 & 5.1$\times$10$^{-7}$\\
STD &  2.6$\pm$0.3 & 0.6$\pm$0.04 & 7.8$\pm$0.2 & 23.3$\pm$0.4 & 23.3$\pm$0.4 & 1.9$\pm$0.04 & 6.6 & 2.5$\times$10$^{-7}$\\
\hline
\end{tabular}
\end{threeparttable}
\end{table*}

\begin{table*}
\centering
\caption{DM and total mass of UGC 8490: HI RC}
\label{tab:tb3}
\begin{threeparttable}
\begin{tabular}{p{1.2cm}p{1.2cm}p{1.2cm}p{1.2cm}p{1.2cm}p{1.2cm}p{1.2cm}p{1.2cm}p{1.2cm}}
\hline
DMHs & M$_\mathrm{T}$  & M$_{\mathrm{DM}}$  
& R$_\mathrm{h}$ & $\mathrm{a_1}$ & $\mathrm{a_2}$ & 
$\mathrm{a_3}$ & $\chi^2_\mathrm{r}$ & EDM\\
\hline
NFW &  1.0$\pm$0.08 & 0.8$\pm$0.05 & 82.5$\pm$2.6 & 139.0$\pm$3.0 & 139.0$\pm$3.0 & 11.0$\pm$0.2 & 1.0 & 3.1$\times$10$^{-7}$\\
BKT &  0.28$\pm$0.03 & 0.04$\pm$0.003 & 2.1$\pm$0.07 & 25.7$\pm$0.6 & 25.7$\pm$0.6 & 2.0$\pm$0.05 & 0.98 & 1.5$\times$10$^{-7}$\\
DCN &  0.6$\pm$0.1 & 0.4$\pm$0.07 & 8.9$\pm$0.2 & 29.6$\pm$0.6 & 29.6$\pm$0.6 & 2.4$\pm$0.05 & 1.0 & 1.4$\times$10$^{-6}$\\
EIN &  0.3$\pm$0.05 & 0.1$\pm$0.02 & 34.0$\pm$1.9 & 11.8$\pm$0.4 & 11.8$\pm$0.4 & 0.94$\pm$0.04 & 0.75 & 1.1$\times$10$^{-7}$\\
STD &  1.2$\pm$0.09 & 1.0$\pm$0.06 & 23.5$\pm$0.7 & 29.9$\pm$0.6 & 29.9$\pm$0.6 & 2.4$\pm$0.05 & 0.8 & 5.0$\times$10$^{-9}$\\
\hline
\end{tabular}
\end{threeparttable}
\end{table*}

\begin{table*}
\centering
\caption{DM and total mass of UGC 9753: HI RC}
\label{tab:tb4}
\begin{threeparttable}
\begin{tabular}{p{1.2cm}p{1.2cm}p{1.2cm}p{1.2cm}p{1.2cm}p{1.2cm}p{1.2cm}p{1.2cm}p{1.2cm}}
\hline
DMHs & M$_\mathrm{T}$  & M$_{\mathrm{DM}}$  
& R$_\mathrm{h}$ & $\mathrm{a_1}$ & $\mathrm{a_2}$ & 
$\mathrm{a_3}$ & $\chi^2_\mathrm{r}$ & EDM\\
\hline
NFW &  63.4$\pm$0.8 & 61.4$\pm$0.5 & 500.0$\pm$4.2 & 192.0$\pm$1.8 & 192.0$\pm$1.8 & 15.0$\pm$0.1 & 0.97 & 2.8$\times$10$^{-9}$\\
BKT &  2.5$\pm$0.4 & 0.5$\pm$0.1 & 10.6$\pm$1.1 & 8.9$\pm$0.6 & 8.9$\pm$0.6 & 0.7$\pm$0.05 & 0.72 & 8.2$\times$10$^{-9}$\\
DCN &  12.0$\pm$0.9 & 9.9$\pm$0.6 & 96.0$\pm$2.4 & 35.1$\pm$0.6 & 35.1$\pm$0.6 & 2.8$\pm$0.05 & 0.76 & 6.0$\times$10$^{-9}$\\
EIN &  3.2$\pm$0.6 & 1.2$\pm$0.3 & 39.0$\pm$3.7 & 6.5$\pm$0.4 & 6.5$\pm$0.4 & 0.5$\pm$0.03 & 0.74 & 1.3$\times$10$^{-8}$\\
STD &  6.6$\pm$1.2 & 4.6$\pm$0.9 & 39.0$\pm$3.4 & 10.3$\pm$0.6 & 10.3$\pm$0.6 & 0.82$\pm$0.05 & 0.96 & 2.8$\times$10$^{-10}$\\
\hline
\end{tabular}
\end{threeparttable}
\end{table*}

\begin{figure*}
\centering
\includegraphics[width=1.0\hsize]{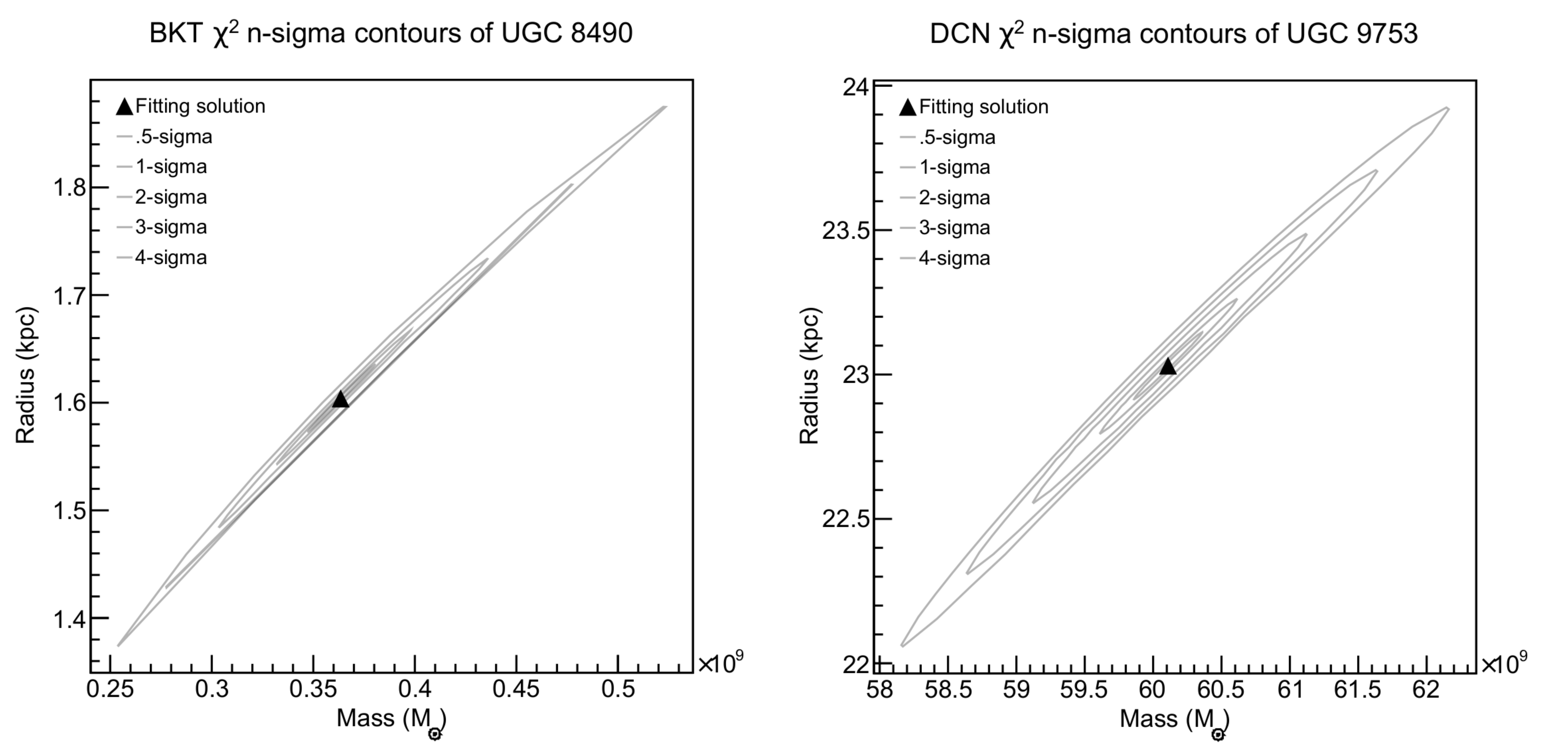}
\caption[f3.pdf]{{\it{Left}}: BKT $\chi^2$ n-sigma contours of the BKT DM fit to the H$\alpha$ DM RC of UGC 8490. The contours
ellipses represent five $\chi^2$ error contour levels. The solution is indicated by a black filled triangle and lies within the 0.5-sigma contour level,
where the optimal solution is expected to stay. {\it{Right}}: Identical information for the DCN $\chi^2$ n-sigma contours of the DCN DM fit to the H$\alpha$ DM RC of UGC 9753.}
~\label{fig3}
\end{figure*}

\begin{figure*}
\centering
\includegraphics[width=1.0\hsize]{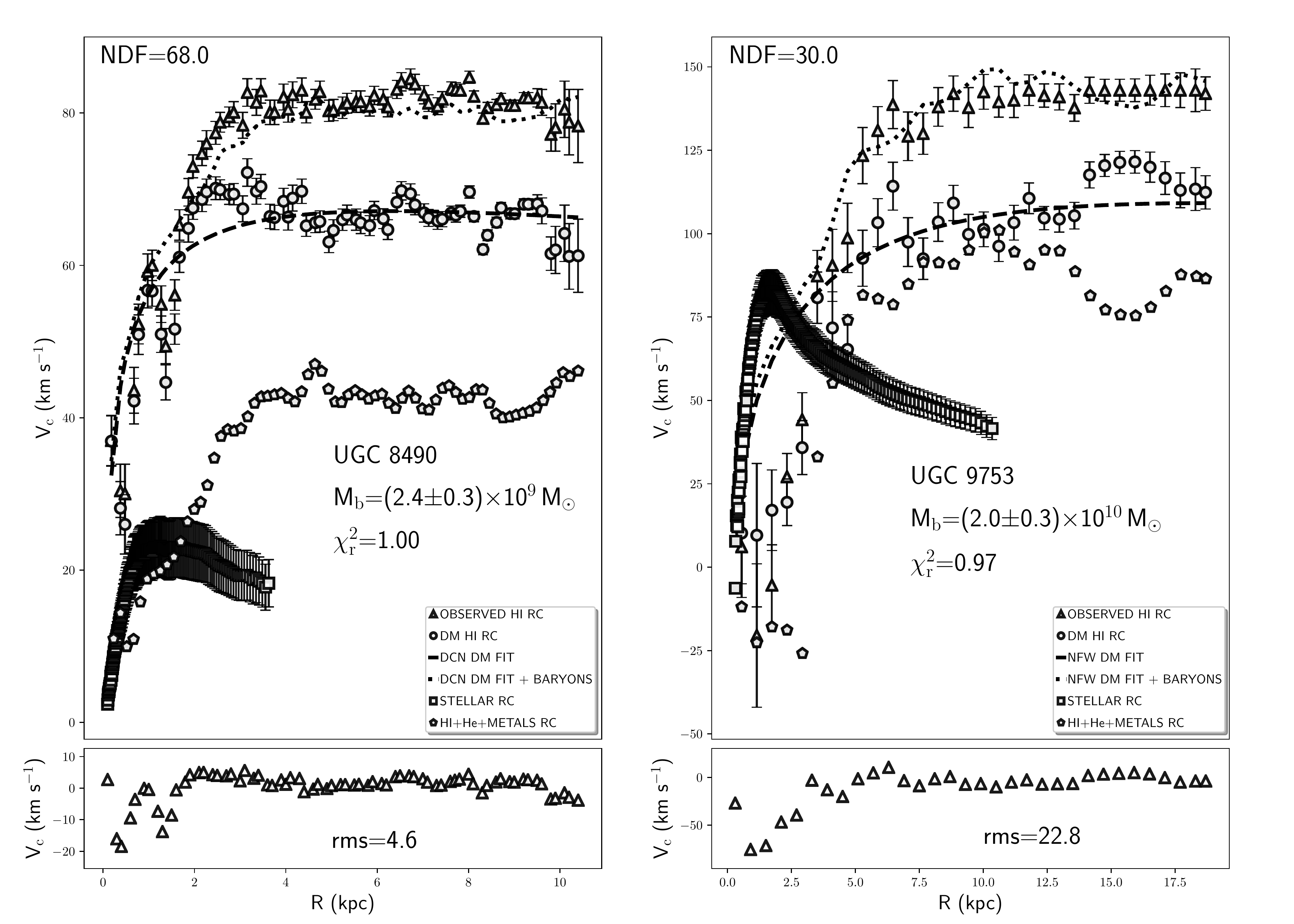}
\caption[f4.pdf]{Identical information of Fig.~\ref{fig2} for the DCN and NFW DM fit to the HI DM RC of UGC 8490 and UGC 9753, respectively.}
~\label{fig4}
\end{figure*}

\begin{figure*}
\centering
\includegraphics[width=1.0\hsize]{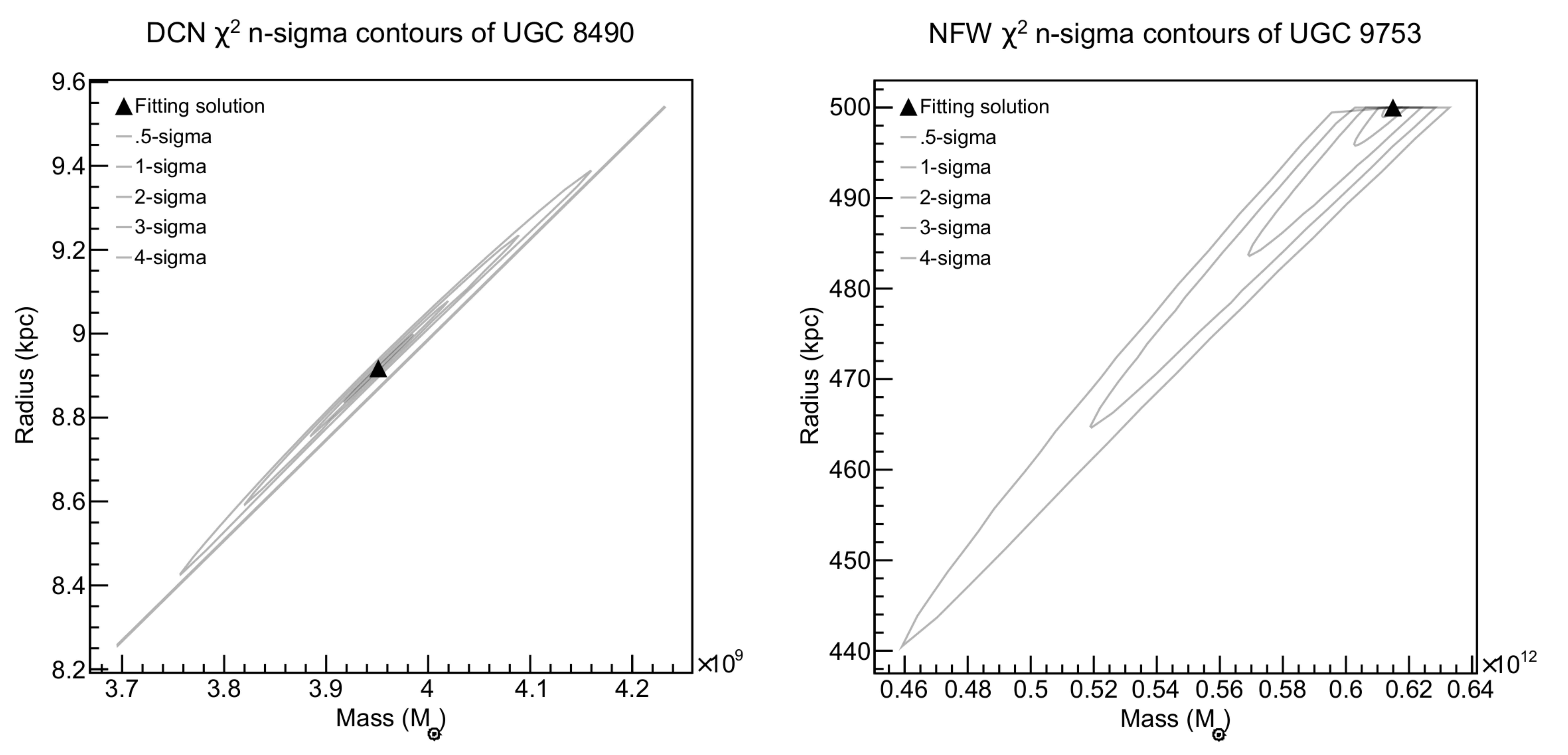}
\caption[f5.pdf]{Identical information of Fig.~\ref{fig3} for the DCN and NFW DM fit to the HI DM RC of UGC 8490 and UGC 9753, respectively.}
~\label{fig5}
\end{figure*}

The principal results for the triaxial spheroidal DM fits to the H$\alpha$ and HI DM RCs of both galaxies are listed in Tables~\ref{tab:tb1},~{\ref{tab:tb2}},~{\ref{tab:tb3}} and~{\ref{tab:tb4}}, separately, and as it is evident from the tabulated $\chi^2_r$, EDM, the DM RCs fits,
and the $\chi^2$ contours, displayed in Figs.~\ref{fig2},~\ref{fig4},~\ref{fig3} and~\ref{fig5}, respectively, the best fitting solutions are those of
cored BKT and cuspy DCN, with an inner slope of 1.3, for the H$\alpha$ DM RCs of UGC 8490 and UGC 9753, respectively. The cuspy DCN of
inner slope 1.2, and the cuspy NFW represent the best fitting solutions for the HI DM RCs of UGC 8490 and UGC 9753. The inward DM distribution of both
galaxies is characterised either by a cored DM density profile, or a cuspy one, and both solutions seem reasonable based on the corresponding DM masses and
scale radii, reported in the respective tables.      

In summary, the results of the current section establish a global oblate spheroidal DM configuration for UGC 8490 and UGC 9753, and a cored
and cuspy inner DM distribution. The validation of the strategy employed to obtain the semi-axes values is performed in the next section through the
application of a standard methodology that computes the resultant torques originated by the encounters among the hypothetical collisionless collective stellar and DM motions and the random displacements of the gaseous components.

\section{Semi-axes determination through the torque method}\label{sec:s7}

In the current section we determine the semi-axes values by means of a widely known alternative methodology, that computes the torques
produced by the dynamical interactions between the gaseous and stellar phases, of \mbox{UGC 8490} and UGC 9753, that move under the gravity of the
corresponding DM haloes. The dynamical interplay among the globally ordered stellar motions and the haphazard gas movements, should delineate the
approximate extension of the DM halo gravitational influence. The corresponding induced perturbation, exerted on the galactic baryonic constituents, should
leave direct evidences of the spatial distribution of the DM halo mass density on the baryonic stellar gravitational potential \citep[e.g.][]{Bekki2002, McQuinn2015, Hu2016}.

\begin{figure*}
\centering
\includegraphics[width=1.0\hsize]{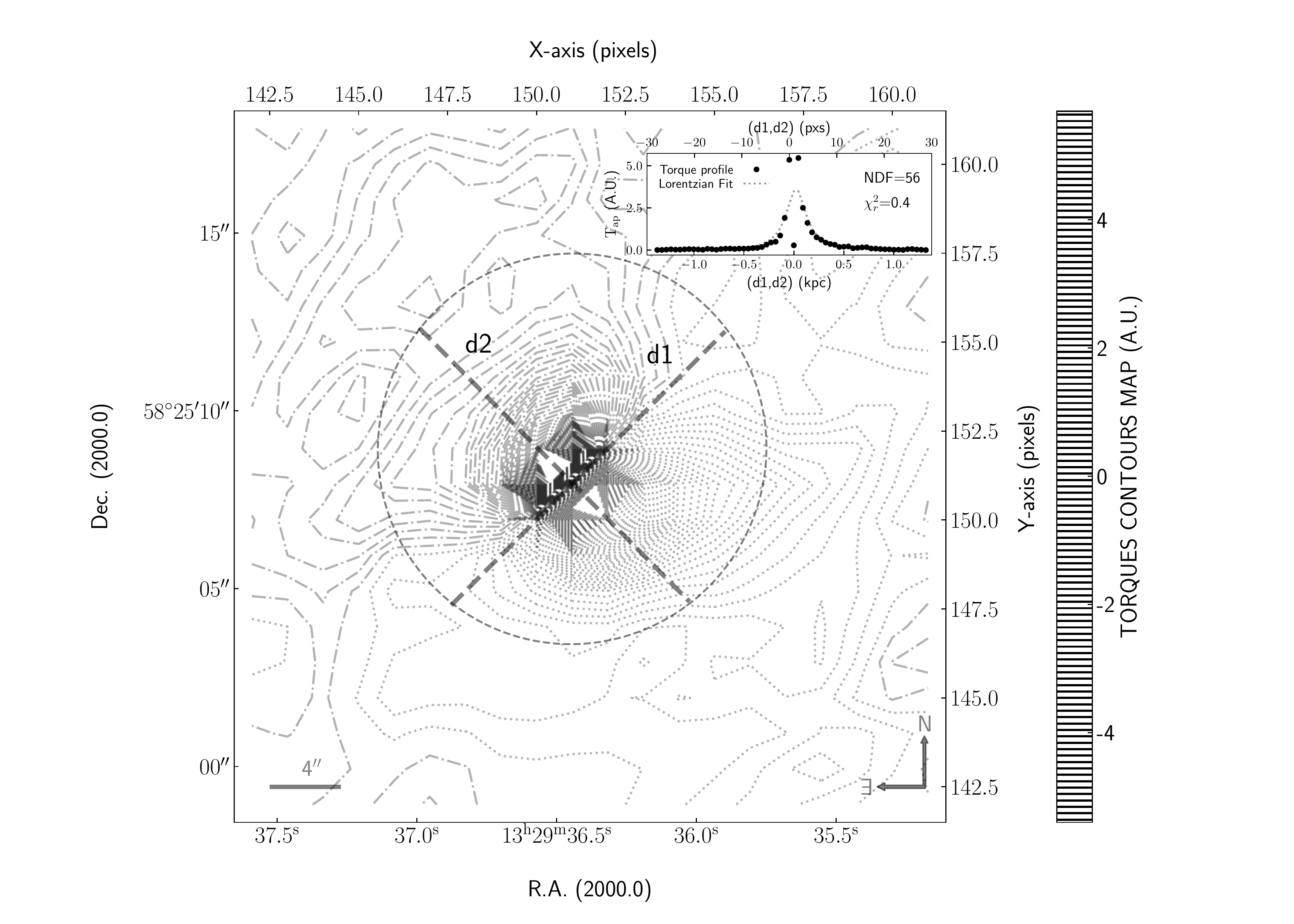}
\caption[f6.pdf]{Torques map contours of UGC 8490, the contours levels range from -5.4 to 5.7 with a constant step of 0.1. The dashed
region encloses the area of maximum perturbation induced by the torques computed at each pixel. The dotted contour lines indicates the zones where the
gravitational torques field is negative, whereas the dash-dotted contour lines denotes the areas where the gravitational torques field is positive. The
inserted panel shows the Lorentzian fit to the profile that represents the average of the two profiles extracted from the Torques map along the directions
$\mathrm{d1}$ and $\mathrm{d2}$. The value of the Lorentzian peak corresponds to the length of the semi-axis of the triaxial spheroid perpendicular to the
EP.}
~\label{fig6}
\end{figure*}

\begin{figure*}
\centering
\includegraphics[width=1.0\hsize]{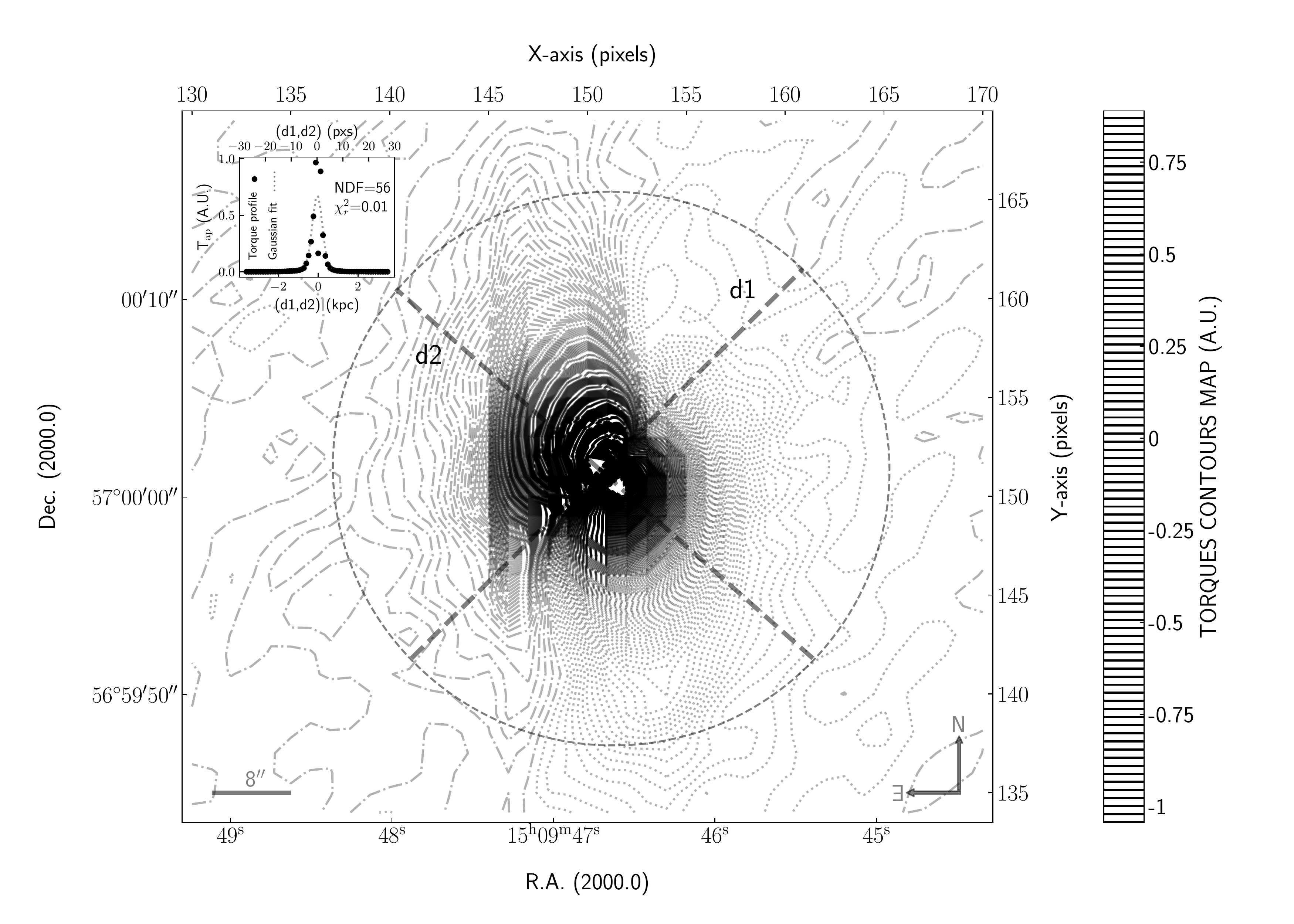}
\caption[f7.pdf]{Identical information of Fig.~\ref{fig6} referred to the torques map of UGC 9753. The contours levels range from -1.04 to
0.9 with a constant step of 0.001. The contours levels interval indicated by the colorbar has a step of 0.022, that approximately corresponds to the central
torque perturbation. The differences among the contours levels steps of the contours map and the corresponding colorbar is necessary, on the one hand, to
depict the entire area of the torque perturbation, and, on the other hand, to obtain a meaningful colorbar.}
~\label{fig7}
\end{figure*}

The specific application of the torques procedure to UGC 8490 and UGC 9753, is accomplished according to the prescription of \citet{Garcia2005}, in particular we compute the gravitational potential maps (GPM) of both galaxies from K band 2MASS \citep{Skrutskie2006} images of UGC 8490 and UGC 9753, through the integral relation of \citet[][Chapter 2, pag. 56, eq. 2.3]{Binney2008}, supposing a M/L of unity, after
proper de-projection, sky subtraction and elimination of any irregularities through appropriate operations of masking and clipping. The photometric de-projection parameters obtained from the HYPERLEDA\footnote{http://leda.univ-lyon1.fr/} astronomical database \citep{Makarov2014}, are position angles and
inclinations of 5.5, 1.5$\degr$ and 58.8, 72.7$\degr$, for UGC 8490 and UGC 9753, respectively. We performed the sky subtraction using the sextractor
package \citep{Bertin1996}. We perform a Fast Fourier Transform of the GPM
of both galaxies, and we determine the corresponding amplitudes, phases, and the central frequencies, to build the Fourier series expansions (FSE) of the
GPM of \mbox{UGC 8490} and UGC 9753. Once we have determined the FSE of the GPM of both galaxies we proceed to obtain the gradients of the latter maps to
ascertain the forces components along the abscissa and the ordinate directions. The torque map of UGC 8490 and \mbox{UGC 9753} is determined by means of the
standard relation, and from those maps we measure the semi-axes of the triaxial spheroidal perturbation. It is worth to notice that the torque component we
have obtained is the one perpendicular to the galactic plane. 
We determine the semi-axes of the torque perturbation, in the EP of UGC 8490 and \mbox{UGC 9753}, from the torques maps of each galaxy, through the IRAF STSDAS Ellipse fitting procedure. The semi-axes perpendicular to the EP of both galaxies, are obtained fitting the vertical profiles of the torques maps of
UGC 8490 and \mbox{UGC 9753}, that originate from the torque perturbation, and represent well discernible features in the torques maps. The corresponding
values of the semi-axes of the torque perturbation of UGC 8490 and UGC 9753 together with their respective uncertainties are $\mathrm{a_1}$=11.0$\pm$0.3,
$\mathrm{a_2}$=10.9$\pm$0.23, and $\mathrm{a_3}$=3.5$\pm$0.12, and, $\mathrm{a_1}$=28.2$\pm$0.8, $\mathrm{a_2}$=27.9$\pm$0.7, and $\mathrm{a_3}$=0.7$\pm$0.06, for \mbox{UGC 8490} and UGC 9753, respectively. The gravitational torques maps of UGC 8490 and UGC 9753 are displayed in figure~(\ref{fig6}) and~(\ref{fig7}),
and the nearly circular area of those figures, highlighted by dashed circles, indicate the regions of the discs of both galaxies where the torque
perturbation is stronger. The average of the semi-axes ratios derived from the fits to the H$\alpha$ and HI DM RCs of \mbox{UGC 8490} and UGC 9753 are
$\mathrm{\frac{a_2}{a_1}}$=1.0, $\mathrm{\frac{a_3}{a_1}}$=0.08, and $\mathrm{\frac{a_3}{a_2}}$=0.08, correspondingly, the average of the semi-axes ratios of the two galaxies, obtained from the torque method, are $\mathrm{\frac{a_2}{a_1}}$=1.0, $\mathrm{\frac{a_3}{a_1}}$=0.17, and $\mathrm{\frac{a_3}{a_2}}$=0.17. The semi-axes values of the DM haloes of UGC 8490 and UGC 9753 describe an oblate DM mass configuration for the two galaxies analysed in this work,
and the corresponding mean semi-axes ratios are approximately concordant with the average semi-axes ratios obtained through the fits to the DM RCs of
UGC 8490 and UGC 9753, and as a consequence the results about the spatial configuration of the DM density of UGC 8490 and \mbox{UGC 9753}, obtained through
the application of the torque method are in general consistent with the corresponding outcomes of the DM fits to the RCs of both galaxies.

The small discrepancy among the semi-axes values obtained through the DM fits and by means of the torque procedure, should not be a matter
of great concern, given that, as explained in the subsequent part of the current section, the two methodologies employed, to obtain the semi-axes values,
analyse very distinct phenomenologies to determine the semi-axes values, and consequently, the results are not expected to be exactly the same. It is
evident that the determination of the DM halo semi-axes through the torque method and by means of the fits to the DM RCs of \mbox{UGC 8490} and UGC 9753, is
based on the analysis of very different phenomena and associated physical processes, given that the DM RC is a measure of the mass density and extent of a
hypothetical virialised density aggregation, whereas the torque method is related to the interaction among different galactic components, and therefore
describes phenomena that are far from any sort of balance among the gravitational forces that determine the dynamical interplay and the consequent evolution
of the dynamics of the distinct galactic constituents, and consequently the outcomes of the two strategies should not necessarily produce similar results.

\subsection{Synthetic data tests: self-generated data and TNG Illustris data}\label{sec:s71}

In the following, we present a short description of the artificial data trials we have performed to evaluate the accurateness of the strategy
devised in this study, whereas, the complete description of those experiments is provided elsewhere. 

Self-generated DM data: we build mock NFW DM RCs considering spherical,
oblate and prolate mass density distributions, and we fit those artificial data employing the rotation formula defined by equation~(\ref{eqn:eq9}), to
determine the robustness of the proposed methodology. We expect that whether the resultant DM masses and scale radii, retrieved by the fitting procedure,
are equal to those settled to build the mock DM RCs data, then the final semi-axes values have to coincide with those adopted to generate the synthetic DM
RCs. The fits to the NFW DM RCs produces some specific values of the DM halo masses, scale radii and the corresponding semi-axes, and, any time that the
resultant DM masses and scale radii are closer or identical to those of the mock data, the correspondent semi-axes are very similar or equal to those of the
synthetic DM RCs. As a further experiment we compute the semi-axes values employing the DM masses and scale radii, originated through the fitting process,
that present greater dissimilarity with respect to those of the artificial DM RCs, originally created, and we obtain semi-axes values that coincide with the
outcomes of the fitting procedure.

TNG Illustris low resolution DM data: we take advantage of the TNG-100-3 DM data \citep{Nelson2019, Pillepich2018, Springel2018, Nelson2018, Naiman2018, Marinacci2018}, to build and fit a spherical DM RC to determine the semi-axes values following the methodology proposed in section~(\ref{sec:s4}) and~(\ref{sec:s5}) of the present manuscript. The complete details about the analysis of the TNG Illustris DM data, the fitting procedure of the determined DM RC,
and the principal results, are given in a separate document, as already stated above. The main outcome of the performed experiment with the TNG Illustris simulation DM data, indicates total concordance with the paramount findings of this study, and the test performed with self-generated DM data. The resulting parameters (in
particular the EIN shape index) of the EIN fit to the TNG Illustris simulated DM RC, that we have exracted and examined in the current tryout, are consistent with the
findings of CH11, specifically their section 6.1, and \citet{Tissera2010}.

The general conclusion is that the semi-axes expressions determined in the present work, represent a very solid
instrument to analyse the DM halo spatial structure, at least from the perspective of a DM RCs fitting analysis. The principal results of this section
constitute additional evidences that support the methodology conceived in this study.

\subsection{Comparison with other works}\label{sec:s72}

In this section we contrast the fitting results for \mbox{UGC 8490} with the outcomes of other studies in the current literature, in particular, with the researches of \citet{Li2018} (L18), and \citet{RT2019} (R19), and in general with other works that perform Cold DM, and Cold DM hydrodynamical simulations, to analyse the different properties of DM halos without and with baryons, especially, the DM halo spatial structure. The comparison for UGC 9753 is not possible, because, in the present literature does not exist any research work, that deals with the DM halo shape, or in general with the content of DM of that galaxy. As anticipated above, the contrast with UGC 8490 is feasible, and for that galaxy, we note a discrepancy
among our M/L values, that vary in the range $\sim$0.2-0.3 and those of L18 and R19, that range in the intervals $\sim$0.75-0.97, and $\sim$0.96-1.4, and, those differences could be determined, for instance, by the different ingredients employed by those authors to build their stellar population synthesis models, such as, for example the selected initial mass function, the Hubble constant, and possibly others. It is important to emphasize that our M/L are in perfect agreement with those of \citet{deBlok2008}, however they are slightly at variance with respect to those of \citet{Oh2008}, likely due to a different choice of the initial mass function, as explained in RP18. The DM masses and scale radii are not listed in the work of L18, therefore a direct comparison with those authors is not feasible, whereas our DM masses and scale radii are distinct from those of R19 (see their table S2), that employs a Self Interactive DM model coupled in the outer galactic parts with a NFW density profile, therefore we argue that the dissimilar modelling conditions, together with the diverse M/L ratio could determine the differences on the DM masses and scale radii. The comparison of our DM masses, scale radii, M/L ratios with those resulting from the work of \citet{Li2020} (L20), indicates higher baryonic disc M/L ratios with respect to our (contrast the M/L
interval of $\sim$0.5-1.24 with the values $\sim$0.2-0.3 of RP18), and consequently the DM masses and scale radii of L20 vary from 2.5$\times$10$^{10}$ to 1.6$\times$10$^{11}$ M$_{\odot}$, and from 0.5 to 11.1 kpc, separately, and are definitively different with respect to the values reported in table~\ref{tab:tb1}, \ref{tab:tb2}, \ref{tab:tb3} and \ref{tab:tb4}, for the DM models of NFW, BKT, DCN and EIN, adopted by us and L20. The DM masses of L20 are higher than our values, and the latter fact seems consistent with the corresponding M/L ratios disparity between the L20 results and our findings. 

The comparison with Cold DM and Cold DM hydrodynamic simulations produces a qualitatively agreement with our results, given that those simulations predict nearly spherical or oblate DM halos, that predict triaxial DM halos at the present epoch \citep[see for instance,][]{Bryan2013, Brainerd2019, Chua2019, Artale2019, Drakos2019}. The value of the ratio among the major and minor axis, determined through our analysis, seems to small when compared with the quoted studies that determine, a range of variation, of that quantity, about $\sim$0.5-0.7. The latter difference could be explained by the kinematic and dynamic peculiarities of the baryonic discs of UGC 8490 and UGC 9753. \citet{Swaters2002} determines, from the HI global profile, that UGC 8490 is strong lopsided, furthermore, \citet{Sicotte1997}, found a strong
kinematic warp in the disc of UGC 8490, from HI observations. According to \citet{Karachentsev2003}, \mbox{UGC 8490} is part of the Canes Venatici I cloud,
and as a consequence, could be affected by environmental phenomena that are typical of galaxy groups, such as, for instance, possible gravitational
encounters with other group members. \mbox{UGC 9753} is an active galaxy, in particular is an intermediate type between a Seyfert 2 and a LINER according to
\citet{Ho1997}, and is classified as a barred galaxy of moderate bar strength (bar index 0.5) by \citet{Buta2015}. \citet{Teo2014} found a very low rate 
($\sim$0.01 M$_{\odot}$ yr$^{-1}$) of HI accretion in the disc of \mbox{UGC 9753}, furthermore, \citet{Sen2006} reported that UGC 9753 belongs to a sparse
group of galaxies, and determined that \mbox{UGC 9753} possesses a dearth of HI of 0.12. The latter study reveals that the HI scarcity detected in UGC 9753, could be caused by tidal ram-pressure stripping, together with HI evaporation produced by kinetically transmitted heat. From the above quoted literature, it
is evident that \mbox{UGC 8490} and UGC 9753, present a very perturbed kinematic and dynamic scenario either within their galactic discs, or, in the surrounding environment, and, all the evidences seem to support the latter view, and, as a consequence, the small axial ratios among their minor and major
axes, of the order of $\sim$0.1, could be connected to those disturbances, given that, according to the study of \citet{Put2019}, the actual shape of
galaxies that have lopsidedness, gas accretion, and other kinematic and dynamic peculiarities, show the corresponding axial ratios of $\approx$0.15, and
even though, the latter result is related to the shape of the whole galaxy, it is well-known, from numerical simulations, that the baryonic and DM
components, can show similar spatial elongations \citep[e.g.][]{Vel2015}. We conclude that our semi-axes ratios seem to be concordant with some author, and
are congruent, with some observed properties of the baryonic discs of \mbox{UGC 8490} and UGC 9753.

\section{Considerations about some of the principal findings}\label{sec:s8}

In this section we discuss the principal results obtained in this work, particularly, the expansion of the most important parts of the methodology of
RP18, that constitutes a fundamental piece of information for the specialisation of the determined gravitational potential relation
to specific mass configurations, and the other important findings about the total and DM mass distribution of UGC 8490 and 
\mbox{UGC 9753}, that constitute an application of the procedural extension of RP18, accomplished in the first part of this study.

The general solution of the Poisson equation obtained in section~(\ref{sec:s3}), supposes that the adopted curvilinear coordinates system is orthogonal (i.e. that the associated metric tensor matrix is diagonal), or equivalently, that can be transformed into orthogonal form, through the diagonalization of the corresponding non diagonal metric tensor matrix. The advantage of this type of solution is that can be utilised to analyse the
dynamics of several different 3D mass density distributions with given volume densities, and for the latter reason represents a very valuable tool to
ascertain which kind of motions are feasible, within those mass density systems, based on the knowledge of the complete force fields of those density aggregations, and consequently of every dynamical and kinematical effects originated. The proposed solution of the Poisson equation determined in this article is also adequate to address dynamic and kinematic problems in 2D or 1D, as the application illustrated in the current work, clearly demonstrate.

It is important to clarify that we fit individually the H$\alpha$ and the HI DM RCs, of UGC 8490 and UGC 9753, because we want to analyse
the DM haloes masses and scale radii, of both galaxies, in a separate manner, for the inner and outer regions of the disc of the analysed galaxies, to
understand the principal differences between the interior and exterior DM mass density configuration of UGC 8490 and UGC 9753. The latter arrangement does
not represent the standard practise \citep[e.g.][]{Swaters2003}, nonetheless, according to us, it should be a more adequate way to study the actual clumpy
structure of the conjectured DM haloes, and to analyse the interaction of the DM halo with distinct parts of the baryonic gravitational potential. We fit the same DM density profiles for the H$\alpha $ and HI DM RCs, and the DM masses and scale radii that result from the fits to the H$\alpha$ and HI DM RCs of
UGC 8490 and \mbox{UGC 9753} are indeed different, probably due to the fact that we measure diverse portions of the DM halo of each galaxy, that interact
with those parts of the baryonic gravitational potential connected either to the ionised or neutral hydrogen. The latter considerations are supported by
several theoretical works that analyse the dynamical and kinematical properties of the DM haloes substructures, some of the most prominent research studies
on that specific subject are, for instance, \citet{Springel2008, Giocoli2010, Lange2018}, among many others.

Although in this work we cannot address the cuspy/core problem, for the reasons clearly stated in the introduction, it is worth to comment something about the DM RCs fitting outcomes. The results of the fits to the DM RCs of UGC 8490 and UGC 9753 establish that both core or cuspy inner slopes are acceptable to explain the inward DM distributions of UGC 8490 and UGC9753, the latter outcome is concordant with some recent works and the general view on the cuspy/core issue that has been developing during many years of research on the subject \citep{Oman2015, Relatores2019, Santos2019}. The current knowledge about the cuspy/core issue seems to indicate that the presence of a cored or cuspy DM density profile is dependent on the dynamical evolutionary state of the baryonic component, along the galaxy formation history, in particular a crucial parameter is the gas density and its relation to the star formation process. The emergent picture suggests that, during the evolution of a galaxy, the core or cuspy behaviour is a transient feature related to the
gas hydrodynamics and its connection to the evolutionary path of the stellar component \citep{Benitez2019, vandokkum2019, Yeh2018}.

The determination of the semi-axes of the supposed DM triaxial spheroid is performed by means of a novel procedure, delineated in section~(\ref{sec:s4}) and appendix~(\ref{sec:a4}), that parametrize the three semi-axes as quotients of the DM haloes masses and scale radii, and the
corresponding volume densities. The fitting procedure to the DM RCs of UGC 8490 and \mbox{UGC 9753}, varies at each fitting iteration the DM masses and
scale radii and consequently generates new values of the three semi-axes, the resultant DM masses and scale radii determine the semi-axes that define the DM
mass density spatial configuration of UGC 8490 and UGC 9753. The results obtained in section~(\ref{sec:s63}) favoured a flattened DM distributions for the
two galaxies analysed in this work, and the successive comparison with the well known strategy of the gravitational torques, to obtain the global
gravitational perturbation induced by the DM haloes on the baryonic components, confirms the findings of the fits to the DM RCs of UGC 8490 and UGC 9753. The methodology devised to obtain the DM haloes semi-axes as parametric ratios of the DM masses, scale radii and the corresponding volume densities, is
substantially different from the analysis of the DM halo triaxiality based on the detection and measurement of the indirect gravitational effects
produced by the DM component on the gaseous and stellar constituents of the galaxies under study, and in the subsequent part of this discussion we
succinctly illustrate the principal differences, among the two strategies, and, at the same time, we highlight the advantages and the drawbacks of both approaches. The recipe ascertained in the current research, to determine the semi-axes of the hypothetical DM triaxial spheroid, estimates the total DM mass
with respect to every axis of symmetry, taking advantage of the invariance of the total DM mass, for any single galaxy examined. The latter procedure does not assume total energy or angular momentum conservation, and does not suppose any particular kinematic degree of freedom, such as for instance rotation, vibration or others, furthermore, all the quantities involved are weighted for the spatial density, that defines the actual DM distribution, that could be completely asymmetric, regardless of the seeming symmetry that is inherent in the semi-axes computation. For all the reasons enumerated, the methodology devised in this work to obtain the semi-axes values of the spheroidal DM distribution of UGC 8490 and UGC 9753, seems adequate to be applied to issues connected with the DM in galaxies.

The comparison with the strategy of the gravitational torques seems satisfactory, nonetheless the exact values of the semi-axes obtained with the methodology devised in the current work do not coincide with those determined through the gravitational torques approach, and one of the possible
reason is that as partially explained at the end of section~(\ref{sec:s7}), they measure the same quantities, through the analysis of very different phenomenological and physical conditions, and therefore we expect only a qualitative concordance of the results. The two procedures determine that UGC 8490 and UGC 9753 present flattened DM distributions, nonetheless the strategy based on the fits of the DM RCs of the two galaxies, produces different semi-axes values for each DM density profile employed, whereas the gravitational torques method gives singular distinct values for each galaxy. The latter difference
is an intrinsic feature of each one of the two approaches, due to the fact that the DM fitting procedure needs a predefined DM model to work, instead the
gravitational torques method does not depend on any DM model. In summary the methodology conceived in this article to obtain the semi-axes of the supposed
DM triaxial spheroid seems a valid and reliable tool, that can be used as a stand-alone reasonable alternative to the gravitational torques strategy, or at
least as a comparative approach, as also confirmed by the synthetic data experiments performed in section~\ref{sec:s71}, and, by the
contrast with the findings of other authors of section~\ref{sec:s72}.

\section{Conclusions}\label{sec:s9}

In the foremost part of this study we extend the methodological approach of RP18, through the diagonalization of a 3SSM without zeros, providing a general solution of the Poisson equation to obtain a relation for the gravitational potential suitable to describe several different mass distributions, and ascertaining relations for the semi-axes of a triaxial spheroid expressed as ratios of the DM models masses and scale radii, and the
corresponding volume densities. In the second part of the present work we apply the results of the first part, to analyse the DM and total mass amount of
UGC 8490 and UGC 9753, in particular to determine the inner and overall DM mass distribution of the two galaxies, by means of the fits to their DM RCs. The
validation of the fitting results is accomplished through the computation of the gravitational torques exercised by the DM haloes of UGC 8490 and UGC 9753 on their respective baryonic constituents, through mock data trials, and the comparison with other research works in the existing literature. In more details the enumeration of the principal findings is the following:

\begin{itemize}

\item The fits to the H$\alpha$ DM RCs of UGC 8490 and \mbox{UGC 9753} establish a BKT cored and DCN cuspy inner DM distribution for
UGC 8490 and UGC 9753, respectively.\\
\item The fits to the HI DM RCs of UGC 8490 and \mbox{UGC 9753} reveal a DCN cuspy and NFW cuspy inner DM configuration for UGC 8490 and UGC 9753, separately.\\
\item The global DM distributions of UGC 8490 and \mbox{UGC 9753} are well fitted by oblate spheroids, and this latter result is confirmed
by the computation of the DM haloes gravitational torques on the corresponding baryonic components of both galaxies, and the estimation of the semi-axes of
the generated perturbations, and also, through the comparison with other studies.

\end{itemize}

The inner cuspy DM distribution of UGC 8490 and UGC 9753, is concordant with recent works on the subject (see the previous section and the
references therein), and indicates the persistence of a long standing problem, still under intense scrutiny, and whose complete solution demands much more observational and theoretical efforts.

\section*{Acknowledgements}

We have made use of the WSRT on the Web Archive. The Westerbork Synthesis Radio Telescope is operated by the Netherlands Institute for Radio Astronomy
ASTRON, with support of NWO. The WHISP observation were carried out with the Westerbork Synthesis Radio Telescope, which is operated by the Netherlands
Foundation for Research in Astronomy (ASTRON) with financial support from the Netherlands Foundation for Scientific Research (NWO). The WHISP project was
carried out at the Kapteyn Astronomical Institute by J. Kamphuis, D. Sijbring and Y. Tang under the supervision of T.S. van Albada, J.M. van der Hulst and
R. Sancisi. This publication makes use of data products from the Two Micron All Sky Survey, which is a joint project of the University of
Massachusetts and the Infrared Processing and Analysis Center/California Institute of Technology, funded by the National Aeronautics and Space
Administration and the National Science Foundation. We acknowledge the usage of the HyperLeda database (http://leda.univ-lyon1.fr). STSDAS and PyRAF are
products of the Space Telescope Science Institute, which is operated by AURA for NASA. IRAF is distributed by the National Optical Astronomy
Observatories, which is operated by the Association of Universities for Research in Astronomy, Inc. (AURA) under cooperative agreement with the National
Science Foundation. We employ the Kapteyn python package \citep{KapteynPackage} to build some of the figures of the present article. This research has made use of the SIMBAD database, operated at CDS, Strasbourg, France \citep{We2000}, "The SIMBAD astronomical database", Wenger et al. This research has made use of the NASA/IPAC Extragalactic Database (NED), which is operated by the Jet Propulsion Laboratory, California Institute of Technology, under contract with the National Aeronautics and
Space Administration. The IllustrisTNG simulations were undertaken with compute time awarded by the Gauss Centre for Supercomputing (GCS) under GCS 
Large-Scale Projects GCS-ILLU and GCS-DWAR on the GCS share of the supercomputer Hazel Hen at the High Performance Computing Center Stuttgart (HLRS), as well as on the
machines of the Max Planck Computing and Data Facility (MPCDF) in Garching, Germany. This research made use of Astropy, a community-developed core Python package for Astronomy \citep{Aspy2018, Aspy2013}. This research made use of matplotlib, a Python library for publication quality graphics \citep{Hunter:2007}. This research made use of NumPy \citep{van2011}. This research made use of SciPy \citep{Virt2020}. The author would like to express his gratitude to Dr. Paola Marziani, for the careful reading of the present manuscript.



\section*{Data Availability}

No new data were generated or analysed in support of this research.




\bibliographystyle{mnras}
\bibliography{biblio} 




\appendix

\section{Diagonalization of a 3SSM without zero elements}\label{sec:a1}

In the present addendum we reduce to diagonal form a 3SSM, whose entries are all different from zero. The latter result
is an essential instrument to orthogonalise systems of curvilinear coordinates that are not orthogonal. Primarily we introduce the matrices 
$\mathrm{M_{ml}}$ and $\mathrm{M_{ll}}$, that represent the 3SSM and the matrix resulting from the diagonalization of the 3SSM, respectively. The
definitions of the 3SSM, the resultant diagonal matrix, their characteristic polynomials and the system of equations generated from the equalities among 
the coefficients of the particular characteristic polynomials of $\mathrm{M_{ml}}$ and $\mathrm{M_{ll}}$ are displayed below.

\begin{align}\label{eqn:ap1}
M_{ml} & = \begin{pmatrix}
v_{11} & v_{12} & v_{13}\\
v_{21} & v_{22} & v_{23}\\
v_{31} & v_{32} & v_{33}\\
\end{pmatrix} & M_{ll} & = 
\begin{pmatrix}
V_{11} & 0 & 0\\
0 & V_{22} & 0\\
0 & 0 & V_{33}\\
\end{pmatrix}
\end{align}

\begin{align}\label{eqn:ap2}
&p1(q)=q^3-q^2(v_{11}+v_{22}+v_{33})+\nonumber\\
&+q(v_{11}v_{22}+v_{11}v_{33}+v_{22}v_{33}-v^2_{12}-v^2_{13}-v^2_{23})+\nonumber\\
&-(v_{11}v_{22}v_{33}+2v_{12}v_{13}v_{23}-v^2_{12}v_{33}-v^2_{13}v_{22}+\nonumber\\
&-v^2_{23}v_{11})=0
\end{align}

\begin{align}\label{eqn:ap3}
&p2(q)=q^3-q^2(V_{11}+V_{22}+V_{33})+\nonumber\\
&+q(V_{11}V_{22}+V_{11}V_{33}+V_{22}V_{33})-V_{11}V_{22}V_{33}=0
\end{align}

\begin{align}\label{eqn:ap4}
&V_{11}+V_{22}+V_{33} = v_{11}+v_{22}+v_{33}\nonumber\\
&V_{11}V_{22}+V_{11}V_{33}+V_{22}V_{33} = v_{11}v_{22}+v_{11}v_{33}+v_{22}v_{33}+\nonumber\\
&-v^2_{12}-v^2_{13}-v^2_{23}\nonumber\\
&V_{11}V_{22}V_{33} = v_{11}v_{22}v_{33}+2v_{12}v_{13}v_{23}-v^2_{12}v_{33}-v^2_{13}v_{22}+\nonumber\\
&-v^2_{23}v_{11}. 
\end{align}

\noindent The system of equations above can be solved considering 9 quadratic equations, that have as unknown quantities the entries of
$\mathrm{M_{ll}}$ arranged according to all possible permutations of their respective indices. In particular, we indicate the diagonal matrix elements as
$\mathrm{V_{ii},V_{jj},V_{kk}}$, where the indices $\mathrm{i,j,k}$ cycle from 1 to 3 according to different combination, and cannot be equal
simultaneously. In detail, the solution strategy equates the quantities $V_{jj}+V_{kk}$ from the first equation of the system above with the
identical binomial sums extracted from the second equation of the same system. The principal hypotheses to apply this procedure are that
$V_{jj}V_{kk}=v_{jj}v_{kk}-v^2_{jk}$, (eq. a), and as a consequence that $V_{jj}V_{ii}+V_{ii}V_{kk}=v_{jj}v_{ii}+v_{ii}v_{kk}-v^2_{ji}-v^2_{ik}$, (eq. b), as we readily verify once obtained the three diagonal matrix terms. This methodology determines three quadratic
equations in the variable $V_{ii}$ that we report below:

\begin{equation}\label{eqn:ap5}
V^2_{ii}-V_{ii}\left[\Tr{(M_{ml})}\right]+v_{jj}v_{ii}+v_{ii}v_{kk}-v^2_{ji}-v^2_{ik}=0.
\end{equation}

\noindent The solutions to equations~(\ref{eqn:ap5}) are the relations~(\ref{eqn:ap6}), that we specify below, together with the corresponding
discriminants:

\begin{align}\label{eqn:ap6}
&V^{(+,-)}_{ii}=\left[\frac{\Tr{(M_{ml})}\pm\Delta^{(i)}}{2}\right]\nonumber\\\nonumber\\
&\Delta^{(i)}=\sqrt{\left[v_{kk}+v_{jj}-v_{ii}\right]^2+4(v^2_{ji}+v^2_{ik})}.
\end{align}

\noindent The remaining diagonal terms $V_{kk}$ and $V_{jj}$ are obtained solving the system of
equations formed by the relations (a) and (b), and, replacing $V_{ii}$ with the relations of equations~(\ref{eqn:ap6}), to determine six quadratic equations, three in the variables $V_{kk}$, and other three in the variable $V_{jj}$. The most important details
are provided below:

\begin{align}\label{eqn:ap7}
&V_{kk}=\left[\Tr{(M_{ml})}-V^{(+,-)}_{ii}-V_{jj}\right] \Rightarrow V_{kk}=V^{(-,+)}_{ii}-V_{jj}\Rightarrow\nonumber\\\nonumber\\
&V^2_{kk}-V_{kk}\left[V^{(-,+)}_{ii}\right]+(v_{jj}v_{kk}-v^2_{jk})=0.
\end{align}

\noindent The exact solutions to the three quadratic equations for $V_{kk}$ are the following:

\begin{align}\label{eqn:ap8}
&V^{(+,-)}_{kk}=\frac{1}{2}\left\{V^{(-,+)}_{ii}\pm\Delta^{(i)}_1\right\}\nonumber\\\nonumber\\
&\Delta^{(i)}_1=\sqrt{\left[V^{(-,+)}_{ii}\right]^2-4(v_{jj}v_{kk}-v^2_{jk})}.
\end{align}

\noindent The three solution for $V_{jj}$ are readily obtained from the sums $V_{jj}+V^{(+,-)}_{kk}$ according to the ensuing description:

\begin{align}\label{eqn:ap9} 
&V^{(-,+)}_{jj}=V^{(-,+)}_{ii}-V^{(+,-)}_{kk}=\frac{1}{2}\left\{V^{(-,+)}_{ii}\mp\Delta^{(i)}_1\right\}.
\end{align}

\noindent The verification of the initial assumptions about the products $V_{jj}V_{kk}$ and the relations $V_{ii}(V_{jj}+V_{kk})$ are provided by the next
expressions:

\begin{align}\label{eqn:ap10}
&V^{(-,+)}_{jj}V^{(+,-)}_{kk}=\frac{1}{4}\left\{\left[V^{(-,+)}_{ii}\right]^2-\left[\Delta^{(i)}_1\right]^2\right\}=\nonumber\\
&=v_{jj}v_{kk}-v^2_{jk}\nonumber\\\nonumber\\
&V^{(+,-)}_{ii}(V^{(-,+)}_{jj}+V^{(+,-)}_{kk})=V^{(+,-)}_{ii}V^{(-,+)}_{ii}=\nonumber\\
&\frac{1}{4}\left\{\left[\Tr{(M_{ml})}\right]^2-\left[\Delta^{(i)}\right]^2\right\}=v_{ii}(v_{jj}+v_{kk})-v^2_{ji}-v^2_{ik}.
\end{align}

\noindent The first two relations of the system expressed by equation~(\ref{eqn:ap4}) could be quickly verified employing the nine
solutions for the diagonal matrix entries $V_{jj}$, $V_{ii}$ and $V_{kk}$. The sum of the fourth and second relation of equation~(\ref{eqn:ap10})
constitutes the second row of the system~(\ref{eqn:ap4}). The first equation of the same system can be obtained immediately through the sum of the
solutions $V^{(-,+)}_{jj}$, $V^{(+,-)}_{ii}$ and $V^{(+,-)}_{kk}$. The third equation is a consequence of the fact that the diagonal matrix is and
endomorphism of the 3SSM and therefore the respective determinants have to be equal. The latter property is identical to the equality between the
characteristic polynomials with null independent variable, as one promptly realises equating equations~(\ref{eqn:ap2}) and~(\ref{eqn:ap3}) with $q=0$.
The procedure to determine some special instances of the nine solutions obtained in the current addendum is delineated at the end of
section~(\ref{sec:s21}).  

\section{Determination of an integral solution of the Poisson equation}\label{sec:a2}

In this addendum we derive an integral solution of the Laplace and Poisson equations, in order to determine a relation of the 3D
gravitational potential as a function of the entries of the diagonal metric tensor matrix. The resultant expression of the gravitational potential is
suitable to represent several different mass density distributions, defined either through orthogonal systems of coordinates, or non orthogonal systems of
coordinates, that can be converted to orthogonal form, by means of the diagonalization of the metric tensor matrix that defines the initial non orthogonal
system of curvilinear coordinates. Primarily we accomplish the solution of the homogeneous part of the Poisson equation, i.e. the Laplace equation. The most
important steps of that result are delineated below. The Laplace equation expressed as a function of the diagonal elements of the metric tensor matrix reads
\citep{Arfken2005}:

\begin{align}\label{eqn:bp1}
&\frac{1}{\sqrt{V}}\left[\sum_{s=1}^{3}\frac{\partial}{\partial u_s}\left(\sqrt{V}V^{ss}\frac{\partial \phi_s(u_1,u_2,u_3)}{\partial u_s}\right)\right]=0
\end{align}

\noindent The three solutions are obtained setting each term in the sum equal to zero, and are detailed below:

\begin{align}\label{eqn:bp2}
&\frac{\partial \phi_{(t,p,q)}(u_1,u_2,u_3)}{\partial u_{(t,p,q)}} =\nonumber\\ 
&=\left[Cs_t \sqrt{\frac{V_{tt}}{V_{pp}V_{qq}}},Cs_p \sqrt{\frac{V_{pp}}{V_{tt}V_{qq}}},Cs_q \sqrt{\frac{V_{qq}}{V_{tt}V_{pp}}}\right]
\end{align}

\noindent where the variables $\mathrm{(t,p,q)}$ take the values 1,2,3, respectively, and the quantity $\mathrm{Cs_{(t,p,q)}}$ are
constants whose units have to be determined for each specific instance separately. The non homogeneous equation i.e. the Poisson equation, can be expressed
in the following manner:

\begin{align}\label{eqn:bp3}
&\frac{1}{\sqrt{V}}\left[\sum_{s=1}^{3}\frac{\partial}{\partial u_s}\left(\sqrt{V}V^{ss}\frac{\partial \phi_s(u_1,u_2,u_3)}{\partial u_s}\right)\right]=\nonumber\\
&=4\pi G\left[\sum_{s=1}^{3}\rho_s(u_1,u_2,u_3)\right]
\end{align} 

\noindent The solution strategy expresses the volume densities as the partial derivative of the corresponding mass densities to transform
the Poisson equation in a homogeneous equation that can be solved equating each terms to zero. The volume densities can be related to the mass densities in
the following way:

\begin{align}\label{eqn:bp4}
&\rho_s(u_1,u_2,u_3)=\frac{1}{\sqrt{V}}\frac{\partial^3 M_s(u_1,u_2,u_3)}{\partial u_1 \partial u_2 \partial u_3}
\end{align}
  
\noindent replacing the volume densities in equation~(\ref{eqn:bp3}), by means of the latter expressions, we obtain the ensuing relations:

\begin{align}\label{eqn:bp5}
&\sum_{s=1}^{3}\frac{\partial }{\partial u_s}\left[\sqrt{V}V^{ss}\frac{\partial \phi_s(u_1,u_2,u_3)}{\partial u_s}+\right.\nonumber\\
&\left.\hspace{3.3cm}-4\pi G\frac{\partial^2 M_s(u_1,u_2,u_3)}{\partial u_p \partial u_q}\right]=0
\end{align}

\noindent In the above expression, the term within square brackets, is constant and consequently we obtain the following solutions of the
Poisson equation:

\begin{align}\label{eqn:bp6}
&\frac{\partial \phi_s(u_1,u_2,u_3)}{\partial u_s} = \frac{4\pi G}{\sqrt{V}V^{ss}}\frac{\partial^2 M_s(u_1,u_2,u_3)}{\partial u_p \partial u_q}+\frac{Cs_{(t,p,q)}}{\sqrt{V}V^{ss}}
\end{align}

\noindent The constants $\mathrm{Cs_{(t,p,q)}}$ are identical to those encountered for the Laplace equation, the quantities 
$\mathrm{\left[\sqrt{V}V^{ss}\right]^{-1}}$ are equal to those reported in equation~(\ref{eqn:bp2}), and the term involving the partial derivatives of the
3D mass density function can be expressed as follows:

\begin{align}\label{eqn:bp7}
&\frac{\partial^2 M_s(u_1,u_2,u_3)}{\partial u_p \partial u_q}=\int_0^{u_s} \rho_s(v_1,v_2,v_3)\sqrt{V} dv_s
\end{align}

\noindent introducing equation~(\ref{eqn:bp7}) into equation~(\ref{eqn:bp6}), we obtain the following solutions of the Poisson equation:

\begin{align}\label{eqn:bp8}
&\frac{\partial \phi_s(u_1,u_2,u_3)}{\partial u_s} =\nonumber\\ 
&\frac{4\pi G}{\sqrt{V}V^{ss}}\int_0^{u_s} \rho_s(v_1,v_2,v_3)\sqrt{V} dv_s+\frac{Cs_{(t,p,q)}}{\sqrt{V}V^{ss}}
\end{align}

\noindent The precise definition of the quantities $\mathrm{\left[\sqrt{V}V^{ss}\right]^{-1}}$ is the following:

\begin{align}\label{eqn:bp9}
&\frac{1}{\sqrt{V}V^{ss}}=\left[\sqrt{\frac{V_{tt}}{V_{pp}V_{qq}}},\sqrt{\frac{V_{pp}}{V_{tt}V_{qq}}},\sqrt{\frac{V_{qq}}{V_{tt}V_{pp}}}\right]
\end{align}

\noindent The variables $\mathrm{(t,p,q)}$ are identical to those defined in equation~(\ref{eqn:bp2}), and the solutions expressed by
equation~(\ref{eqn:bp8}), represent the first order partial derivatives of the gravitational potential of a given mass distribution. The volume density
points that constitute that mass distribution are defined through a set of orthogonal curvilinear coordinates $\mathrm{(u_1,u_2,u_2)}$, nonetheless the
solutions~(\ref{eqn:bp8}) can be applied also to non orthogonal system of coordinates, once reduced to orthogonal form by means of the diagonalization of
the metric tensor matrix that describes the original system of non orthogonal curvilinear coordinates. In section~(\ref{sec:s3}), we consider the solution
determined by setting $\mathrm{s=1}$, in equation~(\ref{eqn:bp8}), that specific relation is fundamental to address the particular question analysed in the
current work.

\section{Metric tensor matrix elements for a triaxial spheroidal geometry}\label{sec:a3}

We consider the correspondence between the rectangular coordinates $x_i=(x_1,x_2,x_3)$ and the triaxial spheroidal curvilinear coordinates
$v_i=(v_1,v_2,v_3)$, expressed by the following relations:

\begin{align}\label{eqn:cp1}
&x_1=v_1a_1\cos{v_2}\sin{v_3}\nonumber\\
&x_2=v_1a_2\sin{v_2}\sin{v_3}\nonumber\\
&x_3=v_1a_3\cos{v_3}
\end{align}

\noindent where $v_1=\left[\sum_{i=1}^3 \frac{x^2_i}{a^2_i}\right]^{\frac{1}{2}}$ represents the radius of the triaxial spheroid and $a_1$, $a_2$ and $a_3$
are the semi-axes, parallel to the $x_1$, $x_2$ and $x_3$ axes, respectively. The azimuthal and zenithal coordinates are $v_2$ and $v_3$, separately. The
metric tensor matrix entries are determined from equation~(\ref{eqn:cp1}), their definitions \citep[e.g.][]{Aramanovich1961} are reported in
equation~(\ref{eqn:cp2}), whereas their expressions are described in equation~(\ref{eqn:cp3}):

\begin{align}\label{eqn:cp2}
&v_{jj}=\sum_{l=1}^3 \left[\frac{\partial x_l}{\partial v_j}\right]^2\, j={1,2,3} & v_{mk}=\sum_{l=1}^3 \left[\frac{\partial x_l}{\partial v_m}\frac{\partial x_l}{\partial v_k}\right]  
\end{align}

\noindent where the integers $\mathrm{m}$ and $\mathrm{k}$ assume the values $(1,1,2)$ and $(2,3,3)$ respectively. The relations of the elements of the
metric tensor matrix, in triaxial spheroidal curvilinear coordinates, are reported below: 

\begin{align}\label{eqn:cp3}
&v_{11}=\sin^2{v_3}\left[a^2_1\cos^2{v_2}+a^2_2\sin^2{v_2}\right]+a^2_3\cos^2{v_3}\nonumber\\
&v_{22}=v^2_1\sin^2{v_3}\left[a^2_1\sin^2{v_2}+a^2_2\cos^2{v_2}\right]\nonumber\\
&v_{33}=v^2_1\left\{\cos^2{v_3}\left[a^2_1\cos^2{v_2}+a^2_2\sin^2{v_2}\right]+a^2_3\sin^2{v_3}\right\}\nonumber\\
&v_{12}=v_1\cos{v_2}\sin{v_2}\sin^2{v_3}\left[a^2_2-a^2_1\right]\nonumber\\
&v_{13}=v_1\cos{v_3}\sin{v_3}\left[a^2_1\cos^2{v_2}+a^2_2\sin^2{v_2}-a^2_3\right]\nonumber\\
&v_{23}=v^2_1\cos{v_2}\sin{v_2}\cos{v_3}\sin{v_3}\left[a^2_2-a^2_1\right].
\end{align}

\noindent The quantity $\sqrt{V}$ is determined through the determinant of the Jacobian matrix, taking advantage of the equality among the determinant of
the metric tensor matrix and the square of the Jacobian determinant.

\section{Triaxial spheroid semi-axes constraints}\label{sec:a4}

This addendum is dedicated to the derivation of the semi-axes relations of a triaxial spheroidal mass density configuration that it is
supposed to describe appropriately the DM haloes of UGC 8490 and UGC 9753. It is demonstrated that the semi-axes of the hypothesised triaxial spheroid can be
expressed as constant quotients of the resultant DM fitting parameters, such as, for instance, the DM halo masses, scale radii, and the corresponding
spatial densities. First of all we obtain the product of the three semi-axes $\mathrm{a_1a_2a_3}$, from the definition of the radial mass density, of an
hypothetical triaxial spheroid:

\begin{equation}\label{eqn:dp1}
a_1a_2a_3 = \frac{1}{4\pi v^2_1 \rho_E(v_1)} \frac{dM_E(v_1)}{dv_1}
\end{equation}

\noindent where the quantity $\mathrm{\rho_E(v_1)}$ represents the radial volume density of the triaxial spheroid. In this appendix we
solely perform the detailed derivation of the semi-axes product $\mathrm{a_1a_2}$, that involve the equatorial ellipse, since the determination of
the products, $\mathrm{a_1a_3}$ and $\mathrm{a_2a_3}$, that correspond to the two ellipses perpendicular to the equatorial ellipse, is based on identical
arguments. The product of the semi-axes of the equatorial ellipse $\mathrm{a_1a_2}$ is obtained from the definition of its surface mass density in the
following way:

\begin{align}\label{eqn:dp2}
&\frac{dM_{EE}(s_1)}{ds_1}=2\pi a_1a_2 \Sigma_{EE}(s_1)s_1\quad s_1=v_1\sin{v_3}\Rightarrow\nonumber\\
&\frac{dM_{EE}(s_1)}{dv_1}=2\pi a_1a_2 \Sigma_{EE}(s_1)v_1\sin^2{v_3}
\end{align}

\noindent The equatorial surface density $\mathrm{\Sigma_{EE}(s_1)}$ can be expressed as a function of the triaxial spheroid radial mass
density $\mathrm{M_{E}(v_1)}$, through the surface density definition and equation~(\ref{eqn:dp1}). The resultant expression of the equatorial surface density is the ensuing:

\begin{align}\label{eqn:dp3}
&\Sigma_{EE}(s_1)=2\int_0^{x_{3m}} \rho_E(v_1) dx_3\quad x_3=v_1a_3\cos{v_3}\Rightarrow\nonumber\\
&\Sigma_{EE}(s_1)=\frac{\cos{v_3}}{2\pi a_1a_2}\int_0^{x_{3m}} \frac{1}{v^2_1}\frac{dM_E(v_1)}{dv_1}dv_1
\end{align}

\noindent The quantity $\mathrm{x_{3m}}$ defines the maximum extension of the $\mathrm{x_3}$ coordinates. Introducing
equation~(\ref{eqn:dp3}) into equation~(\ref{eqn:dp2}), we determine the following expression for the first derivative of the equatorial mass surface
density, and for the product $\mathrm{a_1a_2}$:

\begin{align}\label{eqn:dp4}
&\frac{dM_{EE}(s_1)}{dv_1}=v_1\sin^2{v_3}\cos{v_3}\int_0^{x_{3m}}\frac{1}{v^2_1}\frac{dM_E(v_1)}{dv_1}dv_1=\nonumber\\
&=2\pi a_1a_2\Sigma_{EE}(s_1)v_1\sin^2{v_3}\Rightarrow a_1a_2=\frac{1}{4\pi}\left[\frac{I_{3M}}{I_{3d}}\right]\nonumber\\
&I_{3M}=\int_0^{x_{3m}} \frac{1}{v^2_1}\frac{dM_E(v_1)}{dv_1}dv_1\quad I_{3d}=\int_0^{x_{3m}}\rho_E(v_1)dx_3
\end{align}

\noindent The angular part of the semi-axes relations determined in the present addendum, is computed for constant values of the angles
$\mathrm{v_3}$ and $\mathrm{v_2}$, depending on the considered symmetries within the analysed spheroidal mass distribution, and those constant values are
included as part of the integral ratios $\mathrm{\frac{I_{xM}}{I_{xd}}}$, and, the latter observation is valid also for the derivation of the other semi-axes
products. The integrals $\mathrm{I_{3M}}$ and $\mathrm{I_{3d}}$ can be computed analytically, and as a consequence the ratio of the two integrals is a
constant without physical units. We perform the integrals calculation elsewhere, however we do not report that computation here, because of its lengthiness.
The product $\mathrm{a_1a_2}$, as well as, any other semi-axes products can be expressed by means of the results of the DM fitting parameters, such as, for
instance, the DM halo masses, scale radii, and the corresponding volume densities. The determination of the other semi-axes products follows the same line
of reasoning, considering the surface mass densities of the two other ellipses orthogonal to the equatorial ellipses, and that contain the semi axes
$\mathrm{a_2}$, $\mathrm{a_3}$, and $\mathrm{a_1}$, $\mathrm{a_3}$, respectively. The three semi-axes products are given by the following expressions:

\begin{align}\label{eqn:dp5}
&a_1a_2=\frac{1}{4\pi}\left[\frac{I_{3M}}{I_{3d}}\right]\quad a_1a_3=\frac{1}{4\pi}\left[\frac{I_{2M}}{I_{2d}}\right]\nonumber\\
&a_2a_3=\frac{1}{4\pi}\left[\frac{I_{1M}}{I_{1d}}\right]
\end{align}

\noindent The equations~(\ref{eqn:eq6}) and~(\ref{eqn:eq7}) are obtained from the system of products above, and the precise expressions of
the integral ratios, obtained after the integrals computation, are quotients of two masses of different magnitudes to produce constants, whose values are
lesser or greater than unity, or even equal to unity, and without any physical dimensions. We employ the semi-axes relations of section~(\ref{sec:s3}) to
establish the actual mass density configuration of the DM haloes of UGC 8490 and UGC 9753.

\bsp	
\label{lastpage}
\end{document}